\documentclass[apj]{emulateapj}
\usepackage{apjfonts,float,epsfig,comment}
\usepackage{natbib}
\usepackage{color} 
\def\asca{{\it ASCA\/}}

\def\chandra{{\it Chandra\/}}
\def\fermi{{\it Fermi\/}}

\def\suzaku{{\it Suzaku\/}}

\def\einstein{{\it Einstein\/}}
\def\hst{{\it {\it HST}\/}}

\def\rosat{{\it ROSAT\/}}
\def\rxte{{\it RXTE\/}}

\def\sax{{\it BeppoSAX\/}}
\def\swift{{\it Swift\/}}

\def\heao{{\it HEAO 1\/}}
\def\xmm{{\it XMM-Newton\/}}
\def\nustar{{\it NuSTAR\/}}
\def\xray{\hbox{X-ray}}

\def\ltsima{$\; \buildrel < \over \sim \;$}
\def\simlt{\lower.5ex\hbox{\ltsima}}
\def\gtsima{$\; \buildrel > \over \sim \;$}
\def\simgt{\lower.5ex\hbox{\gtsima}}
\def\kms{\ifmmode{~{\rm km~s^{-1}}}\else{~km s$^{-1}$}\fi}
\def\lsim{\lower0.3em\hbox{$\,\buildrel <\over\sim\,$}}
\def\gsim{\lower0.3em\hbox{$\,\buildrel >\over\sim\,$}}
\def\spose#1{\hbox to 0pt{#1\hss}}
\def\lesssim{\mathrel{\spose{\lower 3pt\hbox{$\mathchar"218$}}
     \raise 2.0pt\hbox{$\mathchar"13C$}}}
\def\gtrsim{\mathrel{\spose{\lower 3pt\hbox{$\mathchar"218$}}
     \raise 2.0pt\hbox{$\mathchar"13E$}}}

\def\h2{H$_2$}

\def\lum{ergs~s$^{-1}$}

\def\arcsec{\mbox{$^{\prime\prime}$}}
\def\arcmin{\mbox{$^\prime$}}
\def\sfr{$M_{\odot}$ yr$^{-1}$}

\begin{document}

\shortauthors{Wik et al.}
\shorttitle{\nustar, \chandra, and VLBA Observations of NGC~253}

\title{Spatially resolving a starburst galaxy at hard X-ray energies: 
\nustar, \chandra, and VLBA observations of NGC~253}

\author{
D.~R.~Wik,\altaffilmark{1,2}
B.~D.~Lehmer,\altaffilmark{1,2}
A.~E.~Hornschemeier,\altaffilmark{1,2}
M.~Yukita,\altaffilmark{1,2}
A.~Ptak,\altaffilmark{1,2}
A.~Zezas,\altaffilmark{3,4}
V.~Antoniou,\altaffilmark{4}
M.~K.~Argo,\altaffilmark{5, 6}
K.~Bechtol,\altaffilmark{7}
S.~Boggs,\altaffilmark{8}
F.~Christensen,\altaffilmark{9}
W.~Craig,\altaffilmark{8,10}
C.~Hailey,\altaffilmark{11}
F.~Harrison,\altaffilmark{12}
R.~Krivonos,\altaffilmark{8}
T.~J.~Maccarone,\altaffilmark{13}
D.~Stern,\altaffilmark{14}
T.~Venters,\altaffilmark{1}
W.~W.~Zhang\altaffilmark{1}
}

\altaffiltext{1}{NASA Goddard Space Flight Center, Code 662, Greenbelt, MD 20771, USA} 
\altaffiltext{2}{The Johns Hopkins University, Homewood Campus, Baltimore, MD 21218, USA}
\altaffiltext{3}{Physics Department, University of Crete, Heraklion, Greece}
\altaffiltext{4}{Harvard-Smithsonian Center for Astrophysics, 60 Garden Street, 
Cambridge, MA 02138, USA}
\altaffiltext{5}{ASTRON, the Netherlands Institute for Radio Astronomy, Postbus 2, 
7990 AA, Dwingeloo, The Netherlands}
\altaffiltext{6}{Jodrell Bank Centre for Astrophysics,
The University of Manchester, Oxford Rd, Manchester M13 9PL, UK}
\altaffiltext{7}{Kavli Institute for Cosmological Physics, Chicago, IL 60637, USA}
\altaffiltext{8}{U.C. Berkeley Space Sciences Laboratory, Berkeley, CA, USA}
\altaffiltext{9}{National Space Institute, Technical University of Denmark, Copenhagen, DK}
\altaffiltext{10}{Lawrence Livermore National Laboratory, Livermore, CA, USA}
\altaffiltext{11}{Columbia University, New York, NY, USA}
\altaffiltext{12}{Caltech Division of Physics, Mathematics and Astronomy, Pasadena, USA}
\altaffiltext{13}{Department of Physics, Texas Tech University, Lubbock, TX 79409, USA}
\altaffiltext{14}{Jet Propulsion Laboratory,
California Institute of Technology, Pasadena, CA 91109, USA}


\begin{abstract}

Prior to the launch of \nustar, it was not feasible to spatially resolve 
the hard ($E > 10$~keV) emission from galaxies beyond the Local Group.  
The combined \nustar\ dataset, comprised
of three $\sim165$~ks observations, allows spatial characterization of the 
hard X-ray 
emission in 
the galaxy NGC~253
for the first time.
As a follow up to our initial study of its nuclear region, we present the first
results concerning the full galaxy
from simultaneous \nustar, \chandra, and VLBA monitoring of 
the local starburst galaxy NGC~253.  
Above $\sim 10$~keV, nearly all the emission is concentrated within 100\arcsec\ of the
galactic center, produced almost exclusively by three nuclear sources,
an off-nuclear ultraluminous X-ray source (ULX),
and a pulsar candidate that we identify for 
the first time in these observations.
We detect 21 distinct sources in energy bands up to 25 keV, mostly consisting
of intermediate state black hole X-ray binaries.
The global X-ray emission of the galaxy -- dominated by the off-nuclear ULX and
nuclear sources, which are also likely ULXs --
falls steeply (photon index $\gtrsim 3$) above
10 keV, consistent with other \nustar-observed ULXs, and no significant excess above
the background is detected at $E > 40$~keV.  
We report upper limits on diffuse inverse Compton emission for a range of spatial models. 
For the most extended morphologies considered, these hard X-ray constraints 
disfavor a dominant inverse Compton component to explain the $\gamma$-ray 
emission detected with \fermi\ and H.E.S.S.
If NGC~253 is typical of starburst galaxies at higher redshift, their contribution
to the $E > 10$~keV cosmic X-ray background is $< 1$\%.

\end{abstract}


\section{Introduction}
\label{sec:intro}

During reionization, a large fraction of the ionizing radiation in the
Universe may not only be generated by active galactic nuclei (AGN), but also by
other sources in starburst galaxies \citep{FLN+13,MFS13,PMM+14}.
Observing these galaxies at high redshift ($z \sim 4$) may soon be possible
with the upcoming \chandra\ Deep Field 7~Ms survey (P.I.\ Niel Brandt).
However, they will be observed primarily at rest-frame energies above $\sim 5$~keV.
To interpret the integrated X-ray emission from these high-$z$ galaxies,
we rely on understanding their hard band spectra, 
which requires determining the nature of the
constituent sources producing it.

The observational effort to constrain the X-ray spectrum of starburst
galaxies has been underway
since the launch of the first hard X-ray experiments  
\citep{BCO+80}.  
Early attempts included stacking the \heao\ and \einstein\ 
data of a sample of 51 FIR-selected
starburst galaxies \citep{RGP95}.  
Such studies revealed a rather hard X-ray spectral slope (photon index $\Gamma < 2$);
however, statistical constraints at $E > 10$~keV were poor, and possible 
contamination from the instrumental background and/or from confused nearby
sources was problematic.
The types of X-ray binaries (XRBs) dominating at hard energies within starburst galaxies
could drive such a hard slope \citep{PR02}. 
Alternatively, the hard X-ray emission may also be due to
a diffuse population of cosmic-ray electrons inverse Compton (IC) scattering the intense
FIR radiation field within the starburst to X-ray energies.
The exact nature of this emission 
is so far largely unconstrained, which is an important problem to solve considering that
star-forming galaxies are the most numerous X-ray emitting extragalactic
population in the Universe \citep[e.g.,][]{Hor+03,Leh+12}. 

The \nustar\ observatory includes the first focusing X-ray optics that operate in orbit 
above 10~keV \citep{Har+13}, dramatically increasing imaging resolution and 
sensitivity at hard X-ray energies.
For the first time, we are able to distinguish individual binaries and diffuse 
non-thermal emission in starburst galaxies and characterize each component
independently.

NGC~253 is the pilot, deep observation of the \nustar\ starburst survey
program, which also includes simultaneous \nustar\ and \chandra\ observations
of Arp~299 \citep{Pta+14}, M82, M83, NGC~3256, and NGC~3310.   
It is an ideal first target 
since it is one of the nearest starburst galaxies
\citep[3.94~Mpc;][]{Kar+03} and subtends an angular extent
\citep[major-axis 23.8\arcmin;][]{Pen80} comparable to the field of view (FOV)
of \nustar\ ($\approx$$13\arcmin\times13\arcmin$).
Over the last few decades, for this reason, NGC~253 has been a prime
target for \xray\ observatories such as  
\einstein\ \citep[e.g.,][]{FT84}, \rosat\ \citep[e.g.,][]{RPS97, DWH98, VP99,
PVK+00}, \asca\ \citep[e.g.,][]{PSY+97}, \sax\ \citep[e.g.,][]{PMC98, Cap+99},
\xmm\ \citep[e.g.,][]{PRS+01, BP05, BPT+07, BPT+08}, \chandra\
\citep[e.g.,][]{SHW+00, WHS+02, MGF+10, MYT11}, and \suzaku\ 
\citep{MYT11,MYT13}.  

Broadly summarizing, the above studies showed that NGC~253 contains
diverse \xray\ emitting populations throughout the galaxy.  
A thin plasma with temperature of $\sim 0.4$~keV
extends several arcminutes along the plane of the disk, centered around the
nucleus \citep{BPT+07,MYT13}.  
The nucleus itself contains a starburst with a star-formation rate
of $\approx$5~\sfr, roughly 70\% of the rate for the entire
galaxy.  
Emanating from the nuclear starburst is a collimated kpc-scale
outflow (with an X-ray component of $kT \sim 1$~keV), 
extending roughly perpendicular to the galactic disk, which is
limb-brightened in diffuse \xray\ emission \citep[e.g.,][]{SHW+00}.  
Within $\sim 150$~pc of the galactic center, a complex
line structure of Fe-K emission has been resolved into at least three spectral
components from Fe~{\small I} at 6.4~keV, Fe~{\small XXV} at 6.7~keV, and
Fe~{\small XXVI} at 7.0~keV, potentially due to the combination of an obscured
AGN, supernova (SN) remnants, and/or XRBs \citep{MYT13}.  
Point sources in this region include a
heavily obscured ($n_{\rm H} \approx$~[6--10]~$\times 10^{23}$~cm$^{-2}$) AGN candidate
and individual XRBs and the collective emission from sources
within star-forming clouds.

A few dozen \xray\ point sources have been detected across
the disk.
Of particular note are three bright 
point sources within a few arcseconds of each other in the galactic center
and another luminous source
 $\approx$30\arcsec\ to their south, which is most likely a
black-hole (BH) XRB \citep{Leh+13}.
Although these sources were not classified as
ultraluminous X-ray sources (ULXs) by \citet{LB05}, who found only one ULX
at the edge of the optical disk in \rosat\ data, they have since been observed
at qualifying luminosities \citep[$L_{\rm X} \gtrsim 10^{39}$~\lum][]{PRS+01,KP09}.
Two other off-nuclear point sources have also been observed with ULX
luminosities in \xmm\ and/or \chandra\ observations \citep{KP09}.

In addition to X-ray emission from compact objects and thermal gas,
star-forming galaxies are expected to produce diffuse non-thermal X-rays
from relativistic particle populations interacting in the galaxies'
strong FIR radiation fields.
Recently, two of the nearest starburst galaxies, NGC~253 and M82,
have been detected at GeV energies with \fermi\ LAT \citep{Abd+10}
and at TeV energies with H.E.S.S. \citep{Ace+09} and VERITAS \citep{Ver+09},
respectively.
Some fraction of this emission is hadronic, originating from the decay
of neutral pions produced by inelastic collisions of 
cosmic-ray nuclei with interstellar gas.
Most of the remainder is leptonic, involving
interactions between cosmic-ray electrons and interstellar gas
(bremsstrahlung) and radiation fields (IC).
If the ratio of accelerated nuclei to electrons is similar in
starbursts and the Milky Way \citep[with nuclei responsible for $\sim99$\% of the
total cosmic-ray radiated power; e.g.,][]{Str+10}, 
it is generally expected that the $\gamma$-ray luminosity
of starbursts results mainly from hadronic emission. 
However, 
significant leptonic emission is predicted by some models
\citep[see, e.g.,][]{DT05, RAP10, PA12, LHB12}.
This distinction may
be important for understanding feedback processes in actively
star-forming environments 
\citep[e.g.,][]{BAK+13,JSE+08,SB13,SDR08,UPS+12},
since the inferred non-thermal energy
density in both cosmic rays and magnetic fields is larger in hadronic scenarios.
At hard X-ray energies, IC emission is the dominant
non-thermal emission process, and its detection
can directly break the degeneracy between the hadronic and leptonic scenarios
because the relevant radiation fields can be estimated from
FIR observations \citep[see, e.g.,][]{LHB12,CF13}.
Upper limits on diffuse IC emission imply lower bounds on
both the cosmic-ray energy density and the strength of magnetic fields.

While the diffuse emission from non-thermal and thermal gas does not vary
over day-to-year timescales, XRB X-ray emission most certainly does.
This variability results from transitions between various accretion states
onto the compact object, during
which a thermal accretion disk and/or non-thermal corona drives the emission
\citep[e.g.,][]{RM06}.
It can also manifest more dramatically in flares, which are often associated with
radio emission: e.g.,
\citet{Gre+72}, who found the first strong flares from Cygnus X-3,
and \citet{TGK+72}, who found the first connection between the X-ray spectral state and 
the radio brightness in Cygnus X-1.  
Most of the well-studied Galactic X-ray transients are low mass XRBs (LMXBs), 
and their radio luminosities are such that current facilities can only find them 
in the very nearest galaxies \citep{Mid+13}.  

However, a few Galactic XRBs have been found to be extremely radio-bright. 
The most radio-luminous is Cygnus X-3, which has flares reaching 
20~Jy \citep[e.g.,][]{MKH10} and is located at a distance of about 9~kpc \citep{PBP+00}.
It is probably not a coincidence that the donor star in Cyg X-3 is a high mass 
Wolf-Rayet star -- it is likely that the jet in Cyg X-3 is far more radiatively efficient than 
other jets because much of the kinetic power is dissipated on a 
small spatial scale through interactions with the stellar wind from the mass donor.  
No such strong flaring has conclusively been seen in other galaxies, 
but there has been the detection of an extremely strong radio flare without an 
X-ray counterpart in M82 \citep{Mux+10,JMF11}, 
which may be the same phenomenon.  
In classical XRBs, the radio emission is well-correlated with the hard X-rays, 
and the radio flares seem to take place at the transition from a 
hard spectral state to a soft one, perhaps due to shocking of the fast moving jet 
against slower-moving older jet material as the jet speeds up \citep{Vad+03}.  
In Cygnus X-3, the situation is slightly different, with strong radio flares being seen 
on the return from the soft state to the hard state \citep{KHM+10}, 
rather than at times of spectral softening like in other systems.  
Given that the very brightest Galactic XRB in the radio shows unusual properties 
relative to other systems, and is clearly associated with a young mass donor, 
searching for more such objects in nearby galaxies with higher star formation rates 
(and thus a higher proportion of high mass XRBs, or HMXBs) 
may help unravel the causes of these differences.

Two previous studies
utilized our nearly simultaneous \nustar\ and \chandra\
observations of NGC~253 to investigate variable sources.
\citet{Leh+13} established that the 3--40 keV X-ray emission 
of the nuclear region is dominated by 
XRB populations and ULX sources rather than accretion onto a supermassive BH.  
\citet{Mac+14} combined the \chandra\ data with archival \chandra\ and \xmm\
observations to reveal another, non-nuclear source 
with dramatic variability.
Its variability is consistent 
with a $\approx$15 hour period, making it a strong new candidate for being a 
rare Wolf-Rayet HMXB.

In this paper, we utilize the \nustar\ and \chandra\ data to investigate the
populations contributing to the galaxy-wide 0.5--30~keV emission from NGC~253.
Our key goals are to
provide the first-ever hard \xray\ spectral constraint on a starburst galaxy by
(1) measuring the accretion states of the bright XRB population 
in a starburst galaxy environment 
and (2) placing the most sensitive constraints on
diffuse IC emission in a nuclear starburst.  
Our paper is organized as follows.
In Section~\ref{sec:obs} we discuss the reduction of the X-ray and radio datasets and the
analysis of the non-\nustar\ observations.
In Section~\ref{sec:cal} we describe the methodology behind the use of calibration products
in the \nustar\ data analysis.
Section~\ref{sec:results} assesses the diffuse and point-like components contributing to the
galaxy-wide emission in the combined observation.
In Section~\ref{sec:var} we investigate the variability of the brightest sources and 
the results of the radio campaign.
Finally, in Section~\ref{sec:disc} we interpret our results and 
discuss future studies.

All \xray\ fluxes and luminosities quoted here have been corrected for
Galactic absorption, assuming the column density in the
direction of NGC~253 of $1.4 \times 10^{20}$~cm$^{-2}$ \citep{SGW92}.
At the distance of NGC~253, 1\arcsec\ subtends a
physical distance of 19~pc.  
Unless stated otherwise, quoted uncertainties
correspond to 90\% confidence intervals.

 \begin{figure}
\plotone{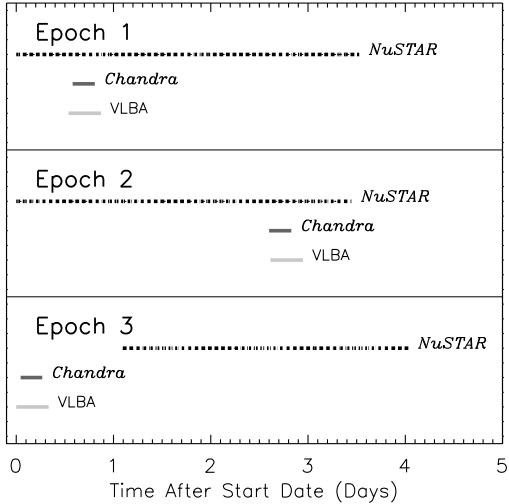}
\caption{Relative \nustar\ (black, dashed lines), \chandra\ (dark gray lines), and VLBA
(light gray lines) observational coverage for each of the three epochs. 
For clarity, we have annotated the total range of observational dates for each epoch. 
The breaks in the \nustar\ observational window are due primarily to 
Earth occultations and passages through the South Atlantic Anomaly.
Additional details are summarized in Table~\ref{tab:obs}.
\label{fig:epochs}}
\end{figure}


\section{Data and Initial Reduction}
\label{sec:obs}

Hard X-ray, soft X-ray, and radio observations were carried out with the
\nustar, \chandra, and VLBA observatories over 3 near simultaneous
epochs, illustrated in Figure~\ref{fig:epochs} and summarized with ObsIDs
in Table~\ref{tab:obs}.
The scientific focus of this paper is based on sources detected in the three
\nustar\ observations, with the \chandra\ data primarily providing identifications.


\subsection{NuSTAR}
\label{sec:obs:nustar}

\begin{figure}
\plotone{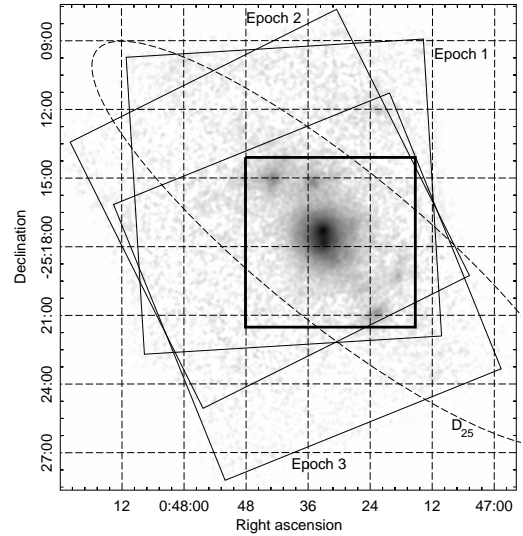}
\caption{The combined, background-subtracted 4-25~keV \nustar\ image 
of NGC~253 from both A and B telescopes
and all three epochs: the approximate 13\arcmin$\times$13\arcmin\, FOV is indicated
in each case.
The image has been smoothed with a 3 pixel ($\sim 7.4\arcsec$) Gaussian
kernel and is logarithmically scaled from 0~cts pix$^{-1}$ to 40~cts pix$^{-1}$.
The dashed ellipse marks the optical $D_{25}$ radius of the disk.
The exposure time of each epoch is given in Table~\ref{tab:obs}.
In this work, we focus on the central overlapping region outlined by the
thickly drawn box.
\label{fig:rawimgs}}
\end{figure}

Each $\approx$165~ks \nustar\ exposure 
utilize data from focal plane modules
``A'' and ``B,'' which image
the same $\approx$$13\arcmin \times 13\arcmin$ region centered on
the nucleus.  
The data were reduced using HEASoft v6.14,
{\tt nustardas}~v1.2.0, and the associated {\tt CALDB} release.  
We began by
bringing level~1 data to level~2 products by running {\tt nupipeline},
which performs a variety of data processing functions, including, e.g.,
filtering out bad pixels, screening for cosmic rays and observational intervals
when the background was too high (e.g., during passes through the South
Atlantic Anomaly), and accurately projecting the events to sky coordinates by
determining the optical axis position and correcting for mast motions.
The task {\tt nupipeline} was executed with the following flags included:
{\tt SAAMODE=STRICT} and {\tt TENTACLE=yes}.
These additional flags reduce the cleaned exposure time by $\sim 15$\%
from what it would otherwise be, but also reduce background uncertainties.
No strong fluctuations are present in light curves produced from the cleaned events,
suggesting a stable background, so no further time periods were excluded.
Images culled from the cleaned events are background subtracted --
following the description in Section~\ref{sec:cal:bgd} --
for each epoch and combined in Figure~\ref{fig:rawimgs}.

\nustar-only source catalogs are not independently created but based on \chandra\
positions (Section~\ref{sec:obs:chandra}) by the methodology described in
Section~\ref{sec:results:ptsrcs:ids}.

\begin{deluxetable}{ccccc}
\tablewidth{0pt}
\tablecaption{Observation Log
\label{tab:obs}}
%
\tablehead{
Observatory  &  Detector & UT Start Date & Observation ID & GTI 
}
 &  & 2012 Sept 1 & 50002031002 & 143.4/143.9~ks \\
\nustar & FPMA/B& 2012 Sept 15 & 50002031004& 141.7/141.5~ks\\
 &  & 2012 Nov 16 & 50002031006& 113.5/113.4~ks \\ \hline
 &   & 2012 Sept 2 & 13830& 19.7~ks\\
\chandra & ACIS-I  & 2012 Sept 18 &13831 &  19.7~ks\\
 &    & 2012 Nov 16 & 13832&  19.2~ks \\ \hline
 & & 2012 Sept 2 & SD679A & 8~hr \\
VLBA & & 2012 Sept 18 & SD679B& 8~hr \\
 & & 2012 Nov 16 & SD679C & 8~hr \\
\end{deluxetable}

   
\subsection{Chandra}
\label{sec:obs:chandra}

All three of the $\approx$20~ks \chandra\ exposures were conducted using single
\hbox{$16.9\arcmin \times 16.9\arcmin$} ACIS-I pointings (ObsIDs 13830, 13831,
and 13832) with the approximate position of the nucleus set as the aimpoint.
For our data reduction, we used {\ttfamily CIAO}~v.~4.4 with {\ttfamily
CALDB}~v.~4.5.0.  We reprocessed our events lists, bringing level~1 to level~2
using the script {\ttfamily chandra\_repro}, which identifies and removes
events from bad pixels and columns, and filters events lists to include only
good time intervals without significant flares and non-cosmic ray events
corresponding to the standard \asca\ grade set (grades 0, 2, 3, 4, 6).  We
constructed an initial \chandra\ source catalog by searching a
\hbox{0.5--7~keV} image using {\ttfamily wavdetect} (run with a point spread
function [PSF] map created using {\ttfamily mkpsfmap}), which was set at a
false-positive probability threshold of $2 \times 10^{-5}$ and run over seven
scales from 1 to 8 (spaced out by factors of $\sqrt{2}$ in wavelet scale: 1,
$\sqrt{2}$, 2, 2$\sqrt{2}$, 4, 4$\sqrt{2}$, and 8).  Each initial \chandra\
source catalog was cross-matched to an equivalent catalog, which we created
following the above procedure using a moderately deep ($\approx$80~ks)
\chandra\ ACIS-S exposure from 2003 September 20 (ObsID: 3931).  The 2003
observation is the deepest \chandra\ image available for NGC~253 and has an
aimpoint close to those of the three 2012 observations.  For the purpose of
comparing point sources in the 2012 observations with those of the deep 2003
exposure (see \citealt{Leh+13}), we chose to register the 2012 aspect solutions
and events lists to the 2003 frame using {\ttfamily CIAO} tools {\ttfamily
reproject\_aspect} and {\ttfamily reproject\_events}, respectively.  The
resulting astrometric reprojections gave very small astrometric adjustments,
including linear translations of \hbox{$\delta x = -0.49$ to +0.37}~pixels and
\hbox{$\delta y = +$0.28 to 0.37}~pixels, rotations of $-0.026$ to
$-0.004$~deg, and pixel scale stretch factors of 0.999963 to 1.000095.  The
final pixel scale of all observations was 0.492~arcsec pix$^{-1}$.

We constructed \chandra\ source catalogs for each of the three epochs in the
\hbox{4--6~keV} bandpass, which overlaps with the \nustar\ response.  
These
catalogs were created by searching \hbox{4--6~keV} images with {\ttfamily
wavdetect} (at a false-positive probability threshold of $10^{-6}$) using a
90\% enclosed count fraction PSF map.  
In Section~\ref{sec:results:ptsrcs:ids}, we utilize the
\hbox{4--6~keV} \chandra\ source catalogs and properties as priors when
computing the \nustar\ point source photometry.


\subsection{VLBA}
\label{sec:obs:vlba}

In order to search for radio emission from \xray\ sources distributed across the
14\arcmin\ field of NGC~253, we made use of the new wide-field capabilities of
the DiFX software correlator \citep{DBP11}, correlating a large number
of sky positions (``phase centers'') in a single correlation pass, thus
allowing us to produce radio maps covering each of the \chandra\ and \nustar\
sources.
This strategy is necessary because very long baseline interferometry (VLBI) 
images made at each phase center 
are typically limited to only a few arcseconds in diameter.
Even though DiFX represents a major gain over standard correlators 
in terms of studying a wider area 
of the galaxy, there is still a limit to the number of correlations one can perform. 
Our strategy was to perform correlations (i.e., search for radio emission) 
at the locations of  \nustar\ and \chandra\ point sources,
which might exhibit rising hard band emission correlated to a radio flare.

We observed NGC~253 at a frequency of 1.4~GHz in three eight-hour sessions,
carried out using all ten
antennas of the VLBA (see Fig.~\ref{fig:epochs} and Table~\ref{tab:obs}).  
At 1.4~GHz, the resolution of our observations is $\sim 5-10$~milli-arcseconds 
(with an elliptical beam due to the low declination of the galaxy) and the largest 
angular scale to which the array is sensitive is $\sim 180$~milli-arcseconds.
Following quick ($\lesssim 24$~hr) processing of the
\chandra\ and \nustar\ images at each epoch, a point-source list was drawn up
of positions to use as correlation phase centers, based on the sources detected
in the \xray\ images.  
Phase centers were also included in a grid covering the
core region where most of the known VLBI-detected components are located.
Correlation parameters were chosen to (1) allow imaging of each field out to a
radius of $\approx 40\arcsec$ with a loss of $\approx 10$\% in sensitivity at the
image edge; (2) allow reliable imaging of fields up to 15\arcmin\ from the
pointing center of the observation; (3) provide a theoretical 5$\sigma$
sensitivity of 150~$\mu$Jy/beam; and (4) keep the correlator output data rate
within practical limits.  
Following correlation, the $\approx 70$ individual
data sets per epoch were transferred to a local machine for processing.  
Data reduction was carried out using standard methods for phase referencing
experiments with the VLBA including: interference rejection, fringe fitting,
and phase and amplitude calibration.  
The first field was calibrated by hand,
then a custom software pipeline was used to transfer the calibration solutions
to each phase center and image the data sets.  
Each field was imaged in four
overlapping squares, each covering a quarter of the entire $\approx 40\arcsec$
field.  
The images were searched for sources with a source finder and inspected
by eye.
Most phase center positions were correlated in more than one epoch; 
these matching calibrated datasets were combined in the $u-v$ plane and 
processed to produce images with a lower noise limit.
A more detailed description of the observations and data analysis
methods will be presented in \citet{Arg+14}.


\section{Further \nustar\ Data Processing}
\label{sec:cal}

The NGC~253 X-ray point source population,
fairly well characterized at $E < 8$~keV by \chandra,
is a crowded field for the \nustar\ PSF (see Fig.~\ref{fig:falsecolor}),
which has an 18\arcsec\ full width at half maximum (FWHM) core
and 58\arcsec\ half power diameter \citep{Har+13}.
Even for sources outside the nuclear region, the wings of the PSF of bright ULXs 
in and to the south of the nucleus complicate standard source analysis
\citep[see Fig.~2 of][]{Leh+13}.
Similarly, local annular background regions would be contaminated by
redistributed source emission.
A gradient in \nustar's $E\la 15$ keV background also prevents
spectra extracted from regions far from sources to be simply scaled and
subtracted from source regions \citep[e.g.,][]{Wik+14}.
We describe our approach to the data analysis below.


\subsection{Background Modeling}
\label{sec:cal:bgd}

We characterize the background using the tool {\tt nuskybgd}, which is
described in detail in \citet{Wik+14}.
Briefly, source-free regions are used to determine the components of a 
background model developed from extragalactic survey observations.
Each component has an assumed spectral and spatial structure, so
once the overall normalization of each component -- which can vary from
observation to observation -- is found somewhere within the FOV,
the model can be extrapolated across the FOV.
We extract spectra from four non-source regions in each epoch, 
simultaneously fit them with the background model, and use those
best-fit parameters to create spatial and spectral backgrounds at source
locations.
These regions cover roughly the entire area within the FOV except for
where source emission is present, which largely corresponds to the
thickly drawn box in Figure~\ref{fig:rawimgs}.
We divide the background
into rectangular segments that align with the roll angle of that epoch and
range in solid angle from 10--40 arcmin$^2$.

In addition to the standard ``Aperture'' background component, which accounts
for stray light (i.e., unreflected photons) 
from the cosmic X-ray background (CXB) reaching the detectors
through the aperture stops, very bright CXB sources 1\arcdeg--5\arcdeg\ from
the target can similarly shine directly on the detectors and distort the background shape
and spectrum.
The Seyfert 2 galaxy NGC~235A is 4.2\arcdeg\ away, and its \swift\ BAT
flux \citep{WMR+09} makes it a marginal candidate for contamination.
During the background modeling, we add a component with its hard X-ray
spectrum in each region with free normalization.
We find that inclusion of the new component does not appreciably affect the resulting
background model; the surface brightness of NGC~235A is roughly comparable
to that of the CXB focused by the optics, which accounts for at most 10\% of the
background below 10~keV.


\subsection{PSF Modeling}
\label{sec:cal:psf}

The \nustar\ PSF shape is well calibrated  \citep[see][for details]{Har+13}
as a function of off-axis angle, which
distorts the PSF into a banana-like shape far ($> 3\arcmin$) from the optical
axis.
The distortions are similar in relative magnitude to those seen in \xmm, which are not nearly as
dramatic as those in \chandra.
Additionally, pointing variations cause a given source's off-axis angle 
to wander $\la 1$\arcmin\, over the course of an observation.
While this motion, removed by a metrology system,
is unimportant for the PSF of sources $\la 3$\arcmin\, 
from the optical axis, at larger off-axis angles the PSF shape for a source can
change non-negligibly during an observation.
A few of our sources are this far off-axis, so we create composite PSFs
by combining PSF models (stored in the {\tt CALDB} as images) 
weighted by the time spent at each off-axis angle.

After attempting to fit these PSFs to sources in our observations, we find
that the model PSF core is sharper than what is present in these data.
Simply smoothing the PSF image by 2~pixels ($\sim 5$\arcsec) yields a much
more satisfactory fit, especially in the core.
This additional smearing of the PSF may result from the accumulation of 
pointing reconstruction errors (i.e., jitter)
over these long exposure times.
A jitter of a few arcseconds would be consistent with \nustar's absolute astrometry,
so shifts in the astrometric solution over a long observation seem reasonable.
The PSFs in the {\tt CALDB}, having been calibrated from shorter observations
of bright sources, may not include this effect.
In any case, we find that the smoothed PSFs appear to successfully capture the 
emission from point sources in these data,
which are the deepest \nustar\ observations to clearly image multiple point sources
across a $\sim 6\arcmin$ FOV outside of the Galactic center.
Note that these PSFs include no energy dependence, even though the PSF
does broaden slightly below $\sim 8$~keV.
Energy-dependent
PSFs appear in versions of the {\tt CALDB} after and including v20131007.

\begin{figure*}
\begin{center}
\includegraphics[width=8.5cm]{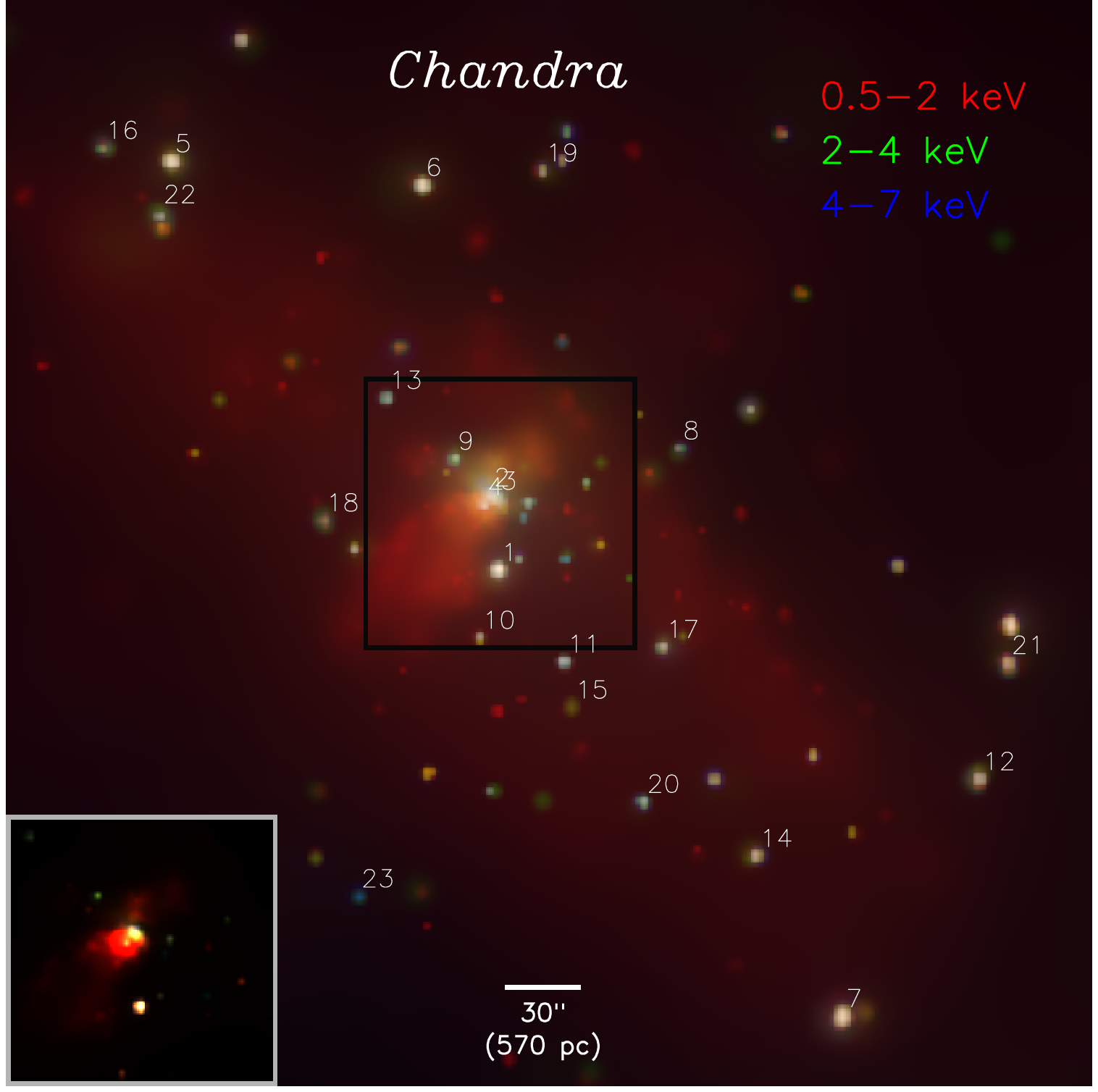}
\includegraphics[width=8.5cm]{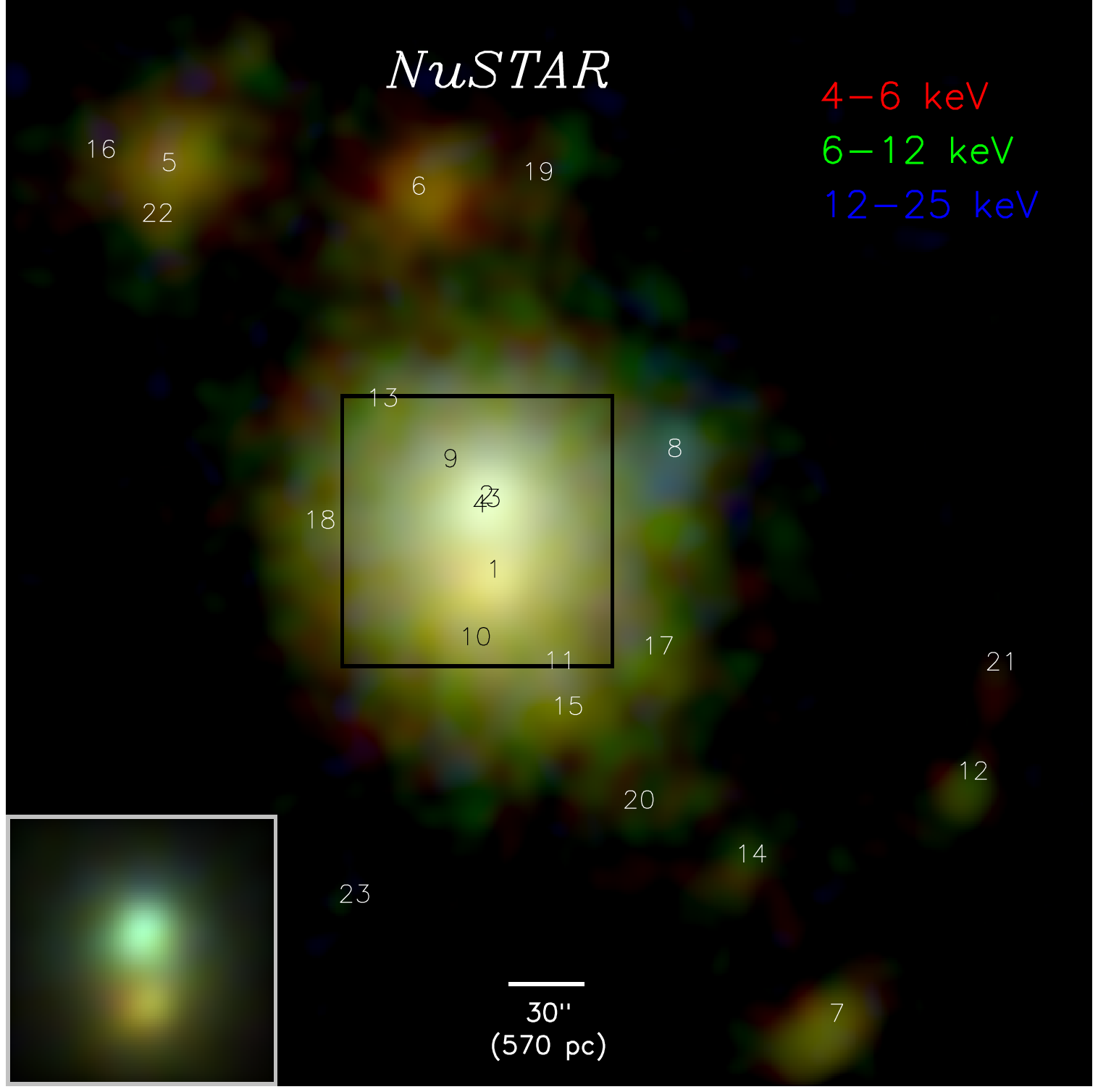}
\caption{False color images (logarithmically scaled)
of the \chandra\ (left) and \nustar\ (right) data in the 7.4\arcmin$\times$7.4\arcmin\
region centered on the nucleus.  
Detected \nustar\ sources are numbered as in Table~\ref{tab:pts}.
Source IDs are sorted by their 4--25~keV count rates in descending order.
We use the higher spatial resolution of \chandra\ to de-convolve the \nustar\ data 
(see Section~\ref{sec:cal:psf} on PSF modeling and 
Section~\ref{sec:results:ptsrcs:ids} on its application), which is particularly important 
in the central regions of the galaxy.
The inset in the lower left corner of each panel, from the central region
outlined in black, shows that part of the image on a linear scale to highlight the
resolving power of \nustar\ (PSF FWHM of $\sim 18\arcsec$).
\label{fig:falsecolor}}
\end{center}
\end{figure*}


\subsection{Exposure Maps and Spectral Responses}
\label{sec:cal:exp}

For off-axis sources, vignetting reduces the overall effective area as a function
of energy, which results in lower exposure times for count rates derived
from images in a given energy band.
To obtain the vignetting function for a particular location on the sky,
we average the functions in the {\tt CALDB}, weighted by the time spent at each
off-axis angle in exactly the same manner as done for the PSF.
Although the effective area declines gradually with off-axis position at energies 
of interest in this paper ($E < 30$~keV), this computation is trivial and produces 
a few percent correction that results in more accurate fluxes.
The vignetting function at a given location is then weighted by a typical source
spectrum, in our case a simple power law with $\Gamma = 2$; 
we use this weighting to prevent the larger amount of higher-energy vignetting 
to unduly influence our results.
Each source now has its own exposure time, corrected such that the count rate
is equivalent to its rate had it been on-axis.

We create spectral response files, RMFs and ARFs, in a similar manner.
For a source extraction region, the composite vignetting function at that location
is multiplied by the on-axis {\tt CALDB} ARF to produce the ARF associated
with that spectrum.
The RMF is detector-based, so we simply use the appropriate {\tt CALDB}
response file modified by an additional absorption associated with that detector.
Although a region may include data from more than one detector, in practice
our regions are dominated by counts from only one detector.


\subsection{Image Fit Methodology and Astrometry Reconstruction}
\label{sec:cal:imfit}

Images are first extracted directly from the cleaned event files in sky coordinates,
individually for each epoch and telescope.
We restrict the FOV of the images to a 181$\times$181
pixel (7.4\arcmin$\times$7.4\arcmin)
box around the nucleus, which contains all the sources associated
with the optical extent of the disk (Fig.~\ref{fig:rawimgs}).
Corresponding to the area of overlap for all three epochs, this sub-image is also
where the total sensitivity and thus signal-to-noise is largest.
Our goal is to combine all six images as accurately as possible.
Because \nustar's absolute astrometry is uncertain to a few arcseconds,
and a small, uncalibrated variable offset between the telescopes still remains,
we must first correct sky positions in the event files before combining the
data of the two telescopes and the three epochs.
This task is made straightforward by the presence of several relatively bright
sources that span the image, all with \chandra\ counterparts that have very
precise positions.
By considering these the true positions, we fit for $x$/$y$-direction shifts and 
rotations that best align the images from the various epochs and telescopes
with these source locations.

We use the same fitting procedure to both get astrometric offsets and measure
source count rates.
Source positions are taken from catalogs of \chandra\ sources, and a PSF
appropriate for that location is created.
A background image is also generated from the
previously derived background model.
The combined background and PSF images serve as a model that is fit
to the images using the Cash statistic \citep{Cash79}, with only
each component's normalization as a free parameter.
We minimize the C-statistic with the Amoeba algorithm \citep{PTV+02},
which is reasonably efficient at avoiding local minima for models without
explicit derivatives.
Because the algorithm completes once a difficult-to-optimize tolerance
parameter is reached, we estimate count rate errors by performing 1000
Monte Carlo realizations of the best-fit model and refitting each one under
the same conditions to ensure that we capture any bias or uncertainty
inherent to the minimization routine.
The normalizations of each component are sorted, and the 90\% uncertainty
is taken as the range that encompasses the inner 900 values.
As long as the uncertainty is dominated by the statistical as opposed to systematic
uncertainties -- excluding those introduced by the fitting algorithm itself --
this method should estimate error ranges accurately.

To obtain the astrometry corrections, we simultaneously fit the A and B
data for a given epoch, in the 4--25~keV band, 
with independent astrometry shifts but linked
source normalizations.
The A and B data are acquired simultaneously themselves, and since they
are calibrated to 3\% \citep{Har+13}, we improve the quality of the fits by
reducing the number of free parameters while introducing negligible
calibration uncertainties.
Although a given source may be at different off-axis angles in the two telescopes,
care is taken to account for differing vignetting in the linking term.
We begin the fitting with only the brightest few sources in the model.
Iteratively, fainter sources are added to the model to ensure the solution is
unbiased by photons from a missing source.
This is necessary because the minimization algorithm will happily skew the
astrometric parameters to better fit positive residuals from a faint, centrally-located
source with the PSF wings from brighter nearby sources.
We consider the astrometric correction to be robust when smoothed residual
images lack large-scale structure and the $x$/$y$ shifts and rotations are
insensitive to minor changes in the fit conditions.
All shifts are $\la 5$\arcsec\ (2 pixels), and the absolute value of rotations are 
$\la 1.5$\arcdeg.
Although the rotations and shifts are small, relative to the PSF FWHM they
are significant and would both blur the combined images and degrade our
ability to fit PSF models to them since the PSF model would be inadequate
for our approach.

To produce combined images,
the sky coordinates in the event files of each epoch and telescope are
adjusted by the astrometric correction before being binned to ensure no
information loss.


\section{Results of Combined Observations}
\label{sec:results}


In Figure~\ref{fig:falsecolor},
false color \chandra\ and \nustar\ images are shown for a
7.4\arcmin$\times$7.4\arcmin\, region centered
on the nucleus of NGC~253.
All results in this section follow from this sub-image, for the simple
reason that we do not detect any sources outside of this region that
also fall within the optical $D_{25}$ radius of the disk.
Diffuse thermal emission from the disk and wind clearly extends over the 
\chandra\ image, but with temperatures too low to be detected by \nustar.
The \nustar\ image is almost entirely comprised of point sources, labeled as in
Table~\ref{tab:pts} (see Section~\ref{sec:results:ptsrcs:ids}), 
which correspond to the same sources indicated in the \chandra\ image.
Source IDs are sorted by their 4-25~keV count rates in descending order.
Because the ULXs in or near the nucleus are so luminous, other near-nuclear
sources fall within their bright wings.
Even so, differences in \nustar\ hardness are still apparent.

\begin{figure*}
\begin{center}
\includegraphics[width=12.0cm]{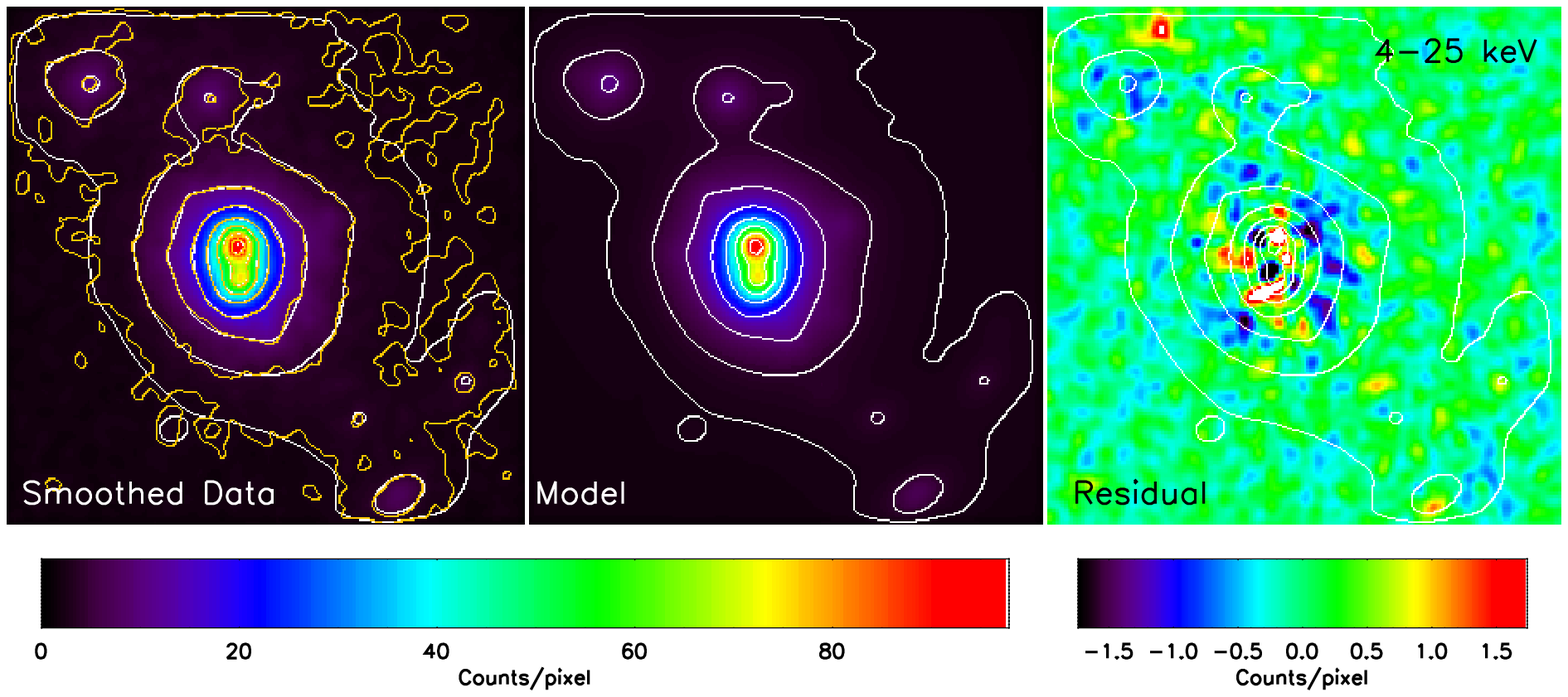}\hspace{0.5in}
\includegraphics[width=12.0cm]{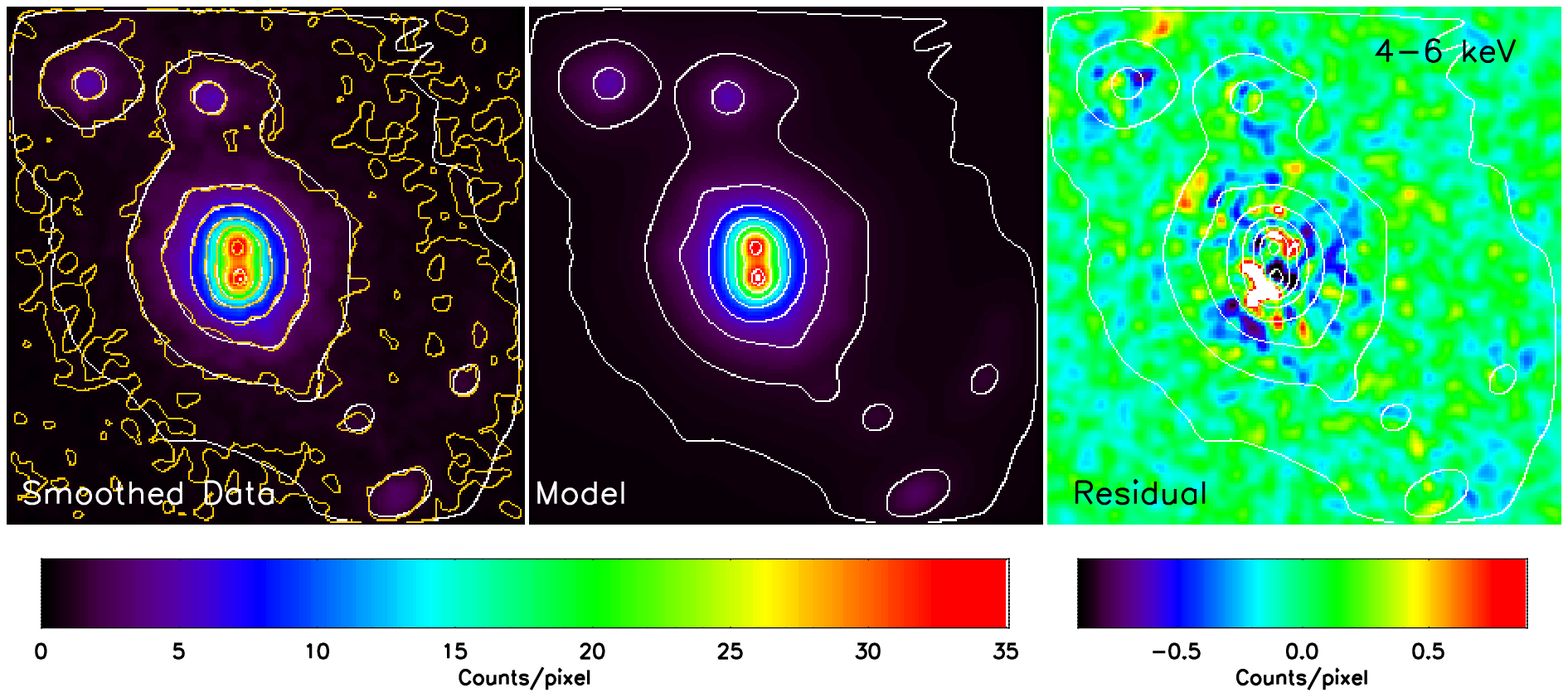}
\includegraphics[width=12.0cm]{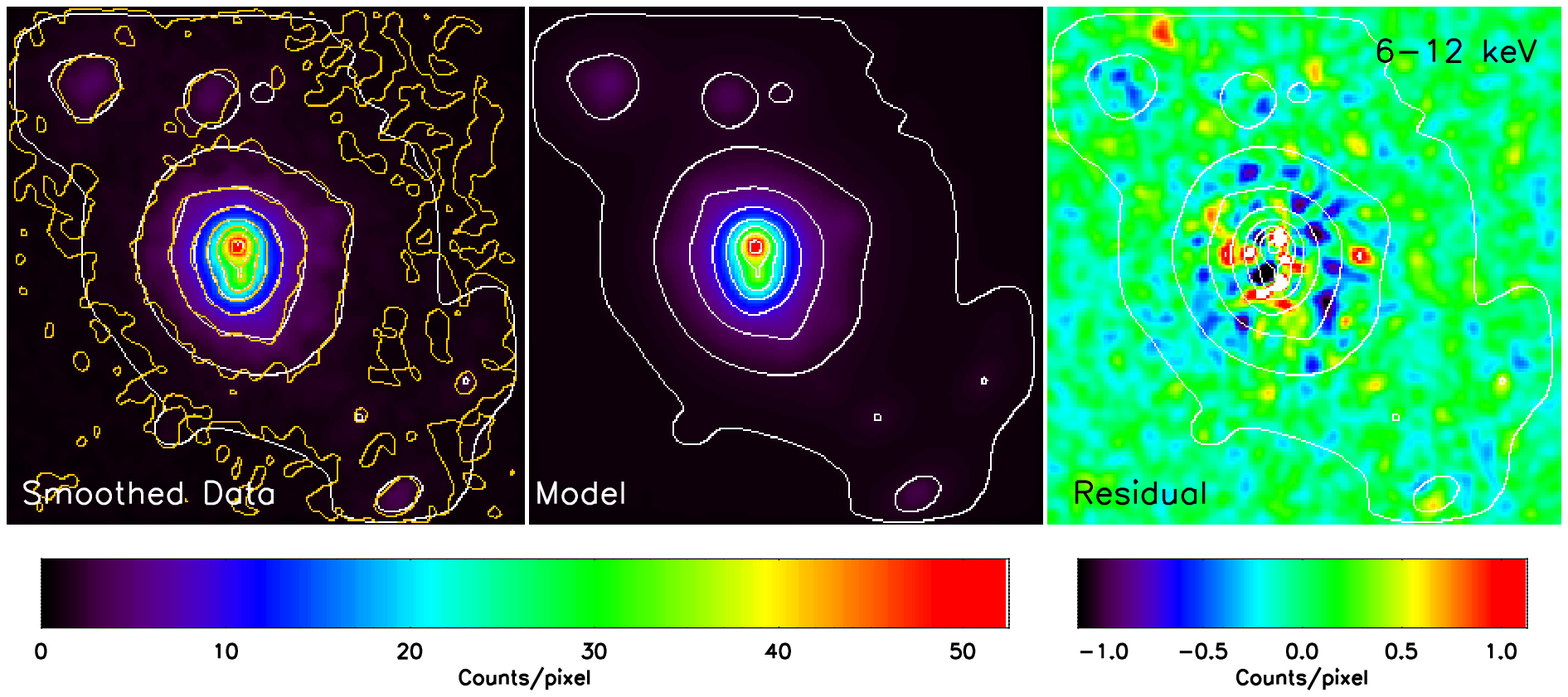}
\includegraphics[width=12.0cm]{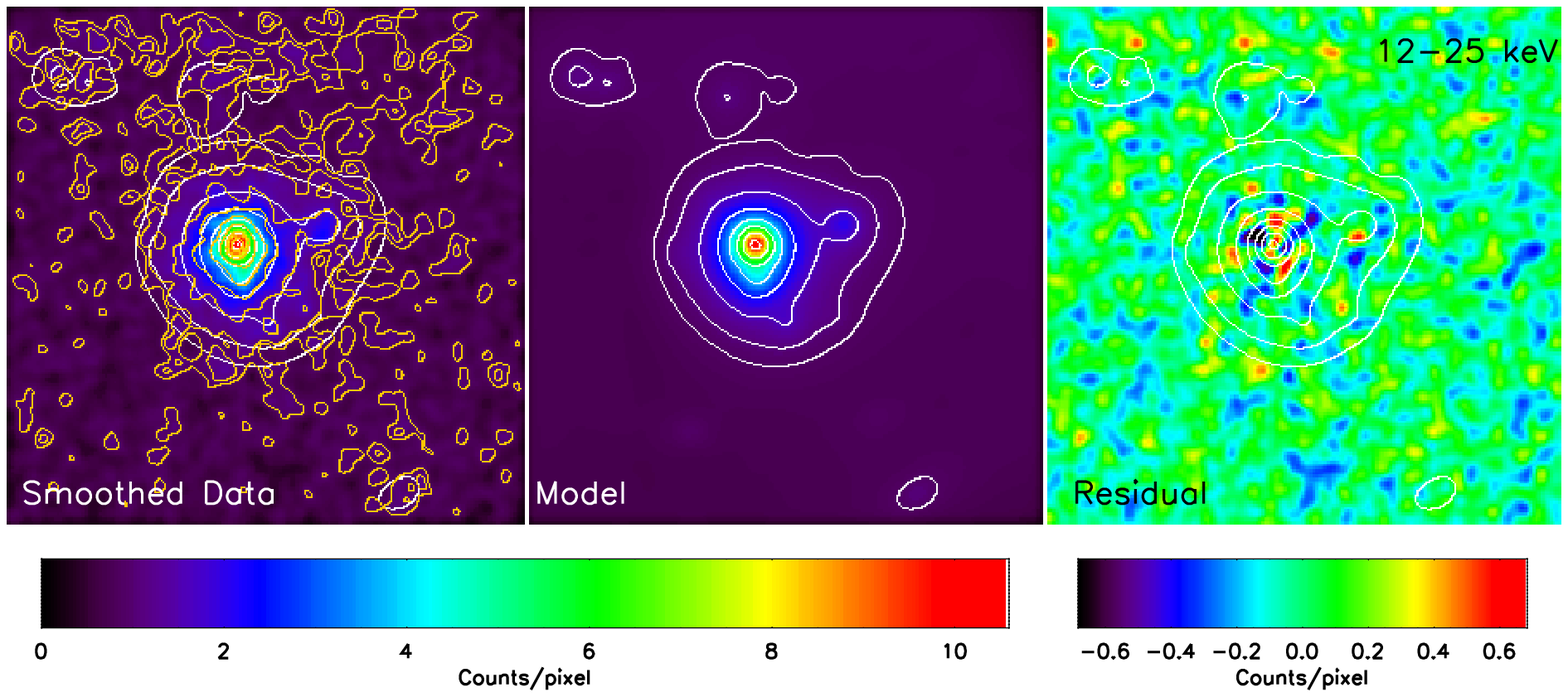}
\caption{
PSF-convolved point source image fits -- 
for the same region shown in Figure~\ref{fig:falsecolor} -- 
in four energy bands, from top to bottom:
4--25~keV, 4--6~keV, 6--12~keV, and 12--25~keV.
In the leftmost panels, the image displays the smoothed, background-subtracted
counts data from all 3 epochs, with the overlaid yellow contours following
the data.
The middle panels show the best-fit model 
(described in Section~\ref{sec:results:ptsrcs:ids})
with the same color scale as the data, and the white contours (also reproduced 
in the left and right panels) follow the underlying smoothed model image.
In the right panels, the residual of the other two panels (data$-$model) is
displayed with its own smaller and more refined color bar.
All images have been smoothed by a Gaussian kernel of 2 pixels
($\sim 5$\arcsec), and the
contours have square-root spacing between the minimum and maximum
values of the model images; both the yellow (data) and model (white) contours
follow identical intensities.
In the left panels, note how well the white contours track the yellow contours,
even where the signal-to-noise is only moderately high,
which is only possible thanks to
the excellent PSF calibration of the \nustar\ telescopes.
The lack of significant structure in the residual images also demonstrates the
success of the fitting process and suggests we have identified all detectable
sources of emission in the central 7.4\arcmin\ (8.5~kpc) of NGC~253.
\label{fig:imgfits}}
\end{center}
\end{figure*}

\subsection{Point Source Properties}
\label{sec:results:ptsrcs}


\subsubsection{Source Identification}
\label{sec:results:ptsrcs:ids}

We assume detectable \nustar\ sources have \chandra\ counterparts, 
since the \chandra\ observations were constructed to exceed the 2--8~keV 
point source sensitivity of the \nustar\ observations.
For point sources in a background-dominated regime, 
spatial resolution is the primary driver of sensitivity, and \chandra's
PSF is more than an order of magnitude smaller than that of \nustar.
While \nustar's larger effective area (a factor of $\sim 2$ at 5~keV) 
allows for a faster accumulation
of source counts, those counts are spread over a much larger detector area,
leading to a similarly high accumulation of background events.
The roughly seven times longer \nustar\ exposure time helps to offset the increase in
noise due to the background, and an isolated source is expected to be detected at
nearly the same significance in these \chandra\ and \nustar\ observations.
However, our sources are not isolated, especially considering the arcminute-scale
wings of the PSF, which complicate the detection of fainter sources near brighter ones.

Initially, sources from the \chandra\ catalogs with the highest 4--6~keV count rates
are included in image fits (see Section~\ref{sec:cal:imfit} for details) 
to the 4--25~keV \nustar\ image.
We inspected the resulting residual \nustar\ images and added
sources from the \chandra\ 2--8~keV band
catalogs where any faint, underlying sources might improve the fit.
In all fits, we also include a spatial model for diffuse thermal gas
based on residual diffuse emission in a 3--7~keV \chandra\ image
(see Section~\ref{sec:results:diffuse:pts} for more details) 
to ensure point source count
rates, especially in the 4--6~keV band, are not biased.
Marginal sources were later removed from our source list
if their rates in three \nustar\ sub-bands were all below a 90\% confidence threshold.
The final rates were found by refitting our four image bands 
(4--6~keV, 6--12~keV, 12--25~keV, and 4--25~keV) with the same culled list
of 23 sources,
which is provided in Table~\ref{tab:pts}.
ID numbers locate each source in Figure~\ref{fig:falsecolor}.

For comparison, in each \chandra\ epoch we detect 36 sources on average
in the 2--7~keV band within our central region of interest.
About two of our 23 sources do not correspond to a \chandra\ source
in a given epoch, although every \nustar\ source has a \chandra\ counterpart in
at least one epoch, as one would expect.
Of the $\sim 15$ sources not detected by \nustar, four are near the brightest
sources 1--4, half of the remaining \chandra\ sources are near to
(and presumably fainter than) other detected
sources, and the rest are more isolated but have the lowest \chandra\ rates.
We therefore detect about two-thirds of the \chandra\ sources in our region of
interest, with the majority of undetected sources missed due to confusion-related issues.

We demonstrate the reliability of the fits to each energy band image in
Figure~\ref{fig:imgfits}.
The varying PSF shape is well modeled across the image (note in particular
Source 7), and the residuals, while not entirely random, indicate that any 
systematic error induced by an erroneously modeled PSF shape is 
$< 5$\% based on the ratio of residual fluctuations (right panels)
over the counts at similar locations (left panels).
The true systematic uncertainty in the PSF shape is probably even smaller
given the impact of statistical fluctuations on the residual images, but since
simple photon statistics dominate our uncertainties, systematic uncertainties
related to the model PSF are not considered further.
While the wings of Sources~1--4 contribute a large fraction of photons to the
locations of surrounding sources, 
the fact that the model (white) contours follow the data (yellow) contours
so well suggests the surrounding source fluxes are not strongly biased.

\begin{figure}
\plotone{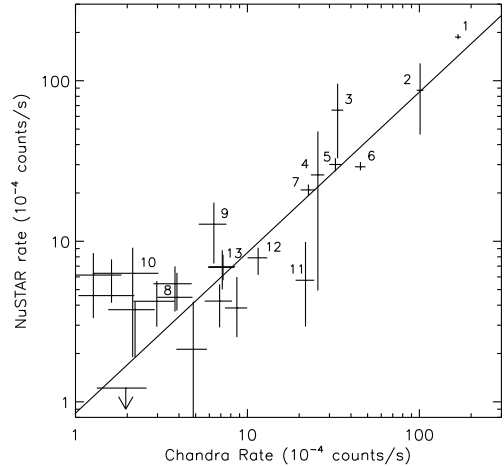}
\caption{
Count rates in the 4--6~keV band for the same sources in the merged \chandra\
and \nustar\ observations; error bars correspond to the 90\% confidence interval.
The diagonal line corresponds to the expected relation ($\sim 0.82$ \nustar\
counts for every \chandra\ count using HEASARC PIMMS), 
and sources are numbered as in Table~\ref{tab:pts}.
The excellent agreement between the two instruments indicates the methods 
outlined in Section~\ref{sec:cal} work well.
\label{fig:ratecompare}}
\end{figure}

As an additional check, we compare the \chandra\ and \nustar\ 4--6~keV rates in
Figure~\ref{fig:ratecompare}.
A source with a power law photon index of 2 should fall on the
solid line, based on a PIMMS count rate conversion (that is rather insensitive
to photon index in any case).
The agreement is good,
although a few sources lie somewhat off the line given their 90\% error bars,
most notably Source 1.
The comparison is done for the combined data 
of all epochs, which are not perfectly simultaneous (Fig.~\ref{fig:epochs}),
primarily because the \nustar\ observations are so much longer.
For Source~1 in particular, the \nustar\ rate increases after the \chandra\ observation
has completed in both Epochs~1 and 3; in Epoch~3 the rate grows monotonically
over the \nustar\ observation.
Additionally, the few faint sources with higher \nustar\ rates (near source 10 on the
plot) may result from variability and/or Eddington bias, since that is near the detection
limit for \nustar.

It should be noted that while the three bright nuclear sources
\citep[Sources 2, 3, and 4, which correspond to sources B, A and C, respectively,
in][]{Leh+13} are individually fit for, their $\sim 3\arcsec$
separations are too small to cleanly separate them spatially with \nustar.
In \citet{Leh+13}, the variability of Source 2 between epochs 
was used to isolate its spectrum in the \nustar\ data; however, we cannot make use of
this fact because all epochs have been combined.
While the 4--6~keV rates seem reasonable for Sources 2--4, given the large errors
on the \nustar\ count rates,
only their summed emission should be considered robust.

\begin{deluxetable*}{rllccc|ccccccc}[t]
\tablewidth{0pt} 
\tablecaption{\nustar\ and \chandra\ Properties of 
\nustar\ NGC~253 Point Sources\tablenotemark{a}
\label{tab:pts}}
\tablehead{ 
 &    &   & & & \colhead{\chandra} & \multicolumn{4}{c}{\nustar\ Count Rates} & &
 \multicolumn{2}{c}{\nustar} \\
 &    &   & \multicolumn{2}{c}{Alt.}  & 
 \colhead{Count Rate} & S & M & H & Full Band & 
$L_X$\tablenotemark{d} &  \multicolumn{2}{c}{Hardness Ratios} \\
 &  \colhead{R.A.}   &  \colhead{Decl.}  & \multicolumn{2}{c}{Name} & 
 \colhead{4--6~keV} & \colhead{4--6~keV} & 
 \colhead{6--12~keV} & \colhead{12--25~keV} & \colhead{4--25~keV} &
 \colhead{4--25~keV} &\underline{(M-S)} & \underline{(H-M)} \\
 \colhead{ID} & \colhead{(J2000)}   & \colhead{(J2000)}  & \tablenotemark{b}  & 
\tablenotemark{c} &  \colhead{(10$^{-4}$ cts s$^{-1}$)} & 
 \colhead{(10$^{-4}$ cts s$^{-1}$)} & \colhead{(10$^{-4}$ cts s$^{-1}$)} & 
 \colhead{(10$^{-4}$ cts s$^{-1}$)} & \colhead{(10$^{-4}$ cts s$^{-1}$)} & 
 ($10^{38}$ erg s$^{-1}$) & (M+S) & (H+M)
}
\startdata
1 & 11.88733 & -25.296933 & X33 & X2 & 167.8 $\pm$ 6.6 & 187.0 $\pm$ \phn6.9 & 153.1 $\pm$ \phn6.9 & \phn10.1 $\pm$ \phn2.5 & 353.2 $\pm$ 10.8 & 20.48 & -0.10$^{+0.03}_{-0.03}$ & -0.88$^{+0.04}_{-0.05}$ \\
2 & 11.88825 & -25.288459 & X34 & X1 & 101.0 $\pm$ 4.6 & \phn87.2 $\pm$ 41.0 & 127.7 $\pm$ 34.5 & \phn47.5 $\pm$ 16.4 & 273.0 $\pm$ 37.1 & 15.83 & \phn0.19$^{+0.24}_{-0.26}$ & -0.46$^{+0.23}_{-0.21}$ \\
3 & 11.88740 & -25.288848 & X34 & X1 & \phn33.5 $\pm$ 2.6 & \phn65.6 $\pm$ 31.5 & \phn99.0 $\pm$ 29.7 & $<$28.5 & 172.0 $\pm$ 32.0 & \phn9.97 & \phn0.20$^{+0.27}_{-0.26}$ & $<$-0.50 \\
4 & 11.88907 & -25.289483 & X34 & X1 & \phn25.7 $\pm$ 2.3 & \phn25.9 $\pm$ 21.7 & \phn68.7 $\pm$ 24.3 & $<$8.2 & \phn90.3 $\pm$ 30.7 & \phn5.23 & \phn0.45$^{+0.36}_{-0.33}$ & $<$-0.64 \\
5 & 11.92817 & -25.250640 & X40 & X6 & \phn32.5 $\pm$ 2.6 & \phn30.1 $\pm$ \phn2.8 & \phn24.0 $\pm$ \phn2.7 & \phn\phn2.0 $\pm$ \phn1.4 & \phn57.8 $\pm$ \phn4.7 & \phn3.35 & -0.11$^{+0.07}_{-0.07}$ & -0.84$^{+0.12}_{-0.12}$ \\
6 & 11.89680 & -25.253328 & X36 & X4 & \phn45.4 $\pm$ 3.2 & \phn29.1 $\pm$ \phn1.7 & \phn16.8 $\pm$ \phn1.6 & \phn\phn1.6 $\pm$ \phn1.1 & \phn48.4 $\pm$ \phn2.8 & \phn2.81 & -0.27$^{+0.05}_{-0.05}$ & -0.83$^{+0.10}_{-0.11}$ \\
7 & 11.84415 & -25.347447 & X21 & X9 & \phn22.6 $\pm$ 2.2 & \phn20.9 $\pm$ \phn1.7 & \phn20.7 $\pm$ \phn1.9 & \phn\phn3.2 $\pm$ \phn1.5 & \phn46.9 $\pm$ \phn3.0 & \phn2.72 & -0.00$^{+0.06}_{-0.06}$ & -0.73$^{+0.10}_{-0.10}$ \\
8 & 11.86456 & -25.283152 &  &  & \phn\phn3.0 $\pm$ 0.8 & \phn\phn4.2 $\pm$ \phn1.4 & \phn11.2 $\pm$ \phn1.6 & \phn\phn7.2 $\pm$ \phn1.4 & \phn22.3 $\pm$ \phn2.6 & \phn1.29 & \phn0.45$^{+0.14}_{-0.14}$ & -0.22$^{+0.11}_{-0.12}$ \\
9 & 11.89275 & -25.284328 &  &  & \phn\phn6.4 $\pm$ 1.2 & \phn12.8 $\pm$ \phn5.1 & \phn\phn7.0 $\pm$ \phn5.3 & $<$4.4 & \phn20.2 $\pm$ \phn8.3 & \phn1.17 & -0.29$^{+0.37}_{-0.37}$ & $<$0.12 \\
10 & 11.88968 & -25.304607 &  &  & \phn\phn2.1 $\pm$ 0.9 & \phn\phn6.3 $\pm$ \phn3.6 & \phn11.1 $\pm$ \phn3.7 & $<$2.6 & \phn19.1 $\pm$ \phn5.4 & \phn1.10 & \phn0.27$^{+0.31}_{-0.30}$ & $<$-0.44 \\
11 & 11.87906 & -25.307325 & T &  & \phn21.8 $\pm$ 2.7 & \phn\phn5.7 $\pm$ \phn3.5 & \phn\phn9.3 $\pm$ \phn3.7 & \phn\phn3.3 $\pm$ \phn1.4 & \phn19.0 $\pm$ \phn5.7 & \phn1.10 & \phn0.24$^{+0.34}_{-0.35}$ & -0.47$^{+0.28}_{-0.36}$ \\
12 & 11.82706 & -25.320597 & X19 & X8 & \phn11.5 $\pm$ 1.5 & \phn\phn7.9 $\pm$ \phn1.5 & \phn\phn8.5 $\pm$ \phn1.6 & $<$0.6 & \phn17.3 $\pm$ \phn2.7 & \phn1.00 & \phn0.04$^{+0.14}_{-0.13}$ & $<$-0.78 \\
13 & 11.90137 & -25.277431 &  &  & \phn\phn7.1 $\pm$ 1.2 & \phn\phn6.9 $\pm$ \phn1.9 & \phn\phn8.7 $\pm$ \phn2.1 & $<$2.2 & \phn16.9 $\pm$ \phn3.4 & \phn0.98 & \phn0.11$^{+0.18}_{-0.18}$ & $<$-0.55 \\
14 & 11.85494 & -25.329321 & X23 & X5 & \phn\phn7.2 $\pm$ 1.3 & \phn\phn6.9 $\pm$ \phn1.3 & \phn\phn7.5 $\pm$ \phn1.5 & $<$2.0 & \phn15.8 $\pm$ \phn2.3 & \phn0.91 & \phn0.04$^{+0.14}_{-0.13}$ & $<$-0.57 \\
15 & 11.87807 & -25.312500 &  &  & \phn\phn1.3 $\pm$ 0.6 & \phn\phn6.2 $\pm$ \phn2.6 & \phn\phn7.5 $\pm$ \phn2.8 & $<$1.0 & \phn13.0 $\pm$ \phn4.2 & \phn0.76 & \phn0.10$^{+0.28}_{-0.28}$ & $<$-0.56 \\
16 & 11.93685 & -25.249135 &  &  & \phn\phn1.6 $\pm$ 0.6 & \phn\phn4.6 $\pm$ \phn1.8 & \phn\phn5.0 $\pm$ \phn1.9 & \phn\phn2.5 $\pm$ \phn1.4 & \phn12.2 $\pm$ \phn3.0 & \phn0.71 & \phn0.04$^{+0.31}_{-0.34}$ & -0.32$^{+0.23}_{-0.42}$ \\
17 & 11.86666 & -25.305651 & X25 &  & \phn\phn8.7 $\pm$ 1.3 & \phn\phn3.8 $\pm$ \phn1.7 & \phn\phn6.6 $\pm$ \phn2.0 & $<$2.2 & \phn11.4 $\pm$ \phn2.9 & \phn0.66 & \phn0.26$^{+0.27}_{-0.25}$ & $<$-0.49 \\
18 & 11.90926 & -25.291312 &  &  & \phn\phn3.8 $\pm$ 1.0 & \phn\phn5.4 $\pm$ \phn1.7 & \phn\phn4.4 $\pm$ \phn1.8 & $<$2.0 & \phn10.7 $\pm$ \phn2.7 & \phn0.62 & -0.11$^{+0.24}_{-0.25}$ & -0.64$^{+0.42}_{-0.35}$ \\
19 & 11.88176 & -25.251690 & X29 &  & \phn\phn2.2 $\pm$ 0.7 & \phn\phn3.7 $\pm$ \phn1.2 & \phn\phn6.1 $\pm$ \phn1.4 & $<$1.5 & \phn10.3 $\pm$ \phn2.2 & \phn0.60 & \phn0.24$^{+0.21}_{-0.20}$ & $<$-0.52 \\
20 & 11.86907 & -25.323165 &  &  & \phn\phn6.9 $\pm$ 1.2 & \phn\phn4.2 $\pm$ \phn1.2 & \phn\phn3.7 $\pm$ \phn1.4 & $<$0.8 & \phn\phn8.0 $\pm$ \phn2.3 & \phn0.47 & -0.06$^{+0.23}_{-0.24}$ & $<$-0.54 \\
21 & 11.82344 & -25.307427 & X18 & X7 & \phn\phn3.9 $\pm$ 0.9 & \phn\phn4.5 $\pm$ \phn1.3 & \phn\phn3.1 $\pm$ \phn1.5 & $<$0.9 & \phn\phn7.0 $\pm$ \phn2.4 & \phn0.41 & -0.18$^{+0.28}_{-0.27}$ & $<$-0.60 \\
22 & 11.92964 & -25.256389 & X42 &  & \phn\phn4.8 $\pm$ 1.0 & \phn\phn2.1 $\pm$ \phn2.1 & \phn\phn4.3 $\pm$ \phn2.2 & $<$0.8 & \phn\phn5.2 $\pm$ \phn3.8 & \phn0.30 & \phn0.34$^{+0.48}_{-0.48}$ & $<$-0.45 \\
23 & 11.90485 & -25.333878 &  &  & \phn\phn2.0 $\pm$ 0.6 & $<$1.2 & $<$3.2 & \phn\phn1.9 $\pm$ \phn1.2 & $<$6.1 & $<$0.36 & $<$1.0 & $>$-0.45

\enddata 
\tablenotetext{a}{Sources' IDs are sorted by their 4-25~keV count rates in descending order.}
\tablenotetext{b}{\citet{VP99, PRS+01}}
\tablenotetext{c}{\citet{LB05}}
\tablenotetext{d}{Simple conversion assuming a typical spectrum-weighted 
effective area across the band of 300 cm$^2$.}
\end{deluxetable*}

\subsubsection{Q-like and Color-Color Diagrams}
\label{sec:results:ptsrcs:cmds}

Hardness-intensity diagrams (also known as ``q'' or ``turtle''-shaped diagrams)
are a simple tool for classifying XRB states.
We create hardness ratios from the rates in Table~\ref{tab:pts} and
compare them to Galactic BH-XRBs in different states
and to Galactic accreting pulsars.
\nustar\ count rates for these Milky Way (MW) sources are derived from
spectral model fits to \rxte\ PCA spectra \citep[for details, see][]{KZ14}.
We have adopted the three-energy band division (over the total range that we detect
emission: $E \lesssim 25$~keV) that provides the best 
discrimination between different types of sources.
Table~\ref{tab:pts} provides these rates for the
soft (S: 4-6~keV), medium (M: 6-12~keV), and hard (H: 12-25~keV) bands.
In Figures~\ref{fig:colorrate} and \ref{fig:colorcolor}, we show the expected locus in the
\nustar\ data for the hard, intermediate, and soft
spectral states of BH XRBs with blue, green and red
squares, respectively. 
We clearly see that both in the ``q''-like and color-color diagrams, 
they follow the well established
pattern from the \rxte\ results \citep[e.g.,][]{RM06, DGK07}.
In addition, we include accreting pulsars (magenta squares), which show
systematically harder spectra, and ULXs from the analysis of \nustar\
data of several sources (orange squares):
\citet[][NGC~1313 X-1 and X-2]{Bac+13};
\citet[][IC~342 X-1 and X2]{Ran+14};
\citet[][Holmberg IX X-1]{Wal+14}; and
\citet[][the ULX in Circinus]{Wal+13}.
Note that the ULX sources appear to have colors similar to intermediate state
Galactic black holes but at much higher luminosities.

The \nustar\ sources from NGC~253 are overplotted on Figures~\ref{fig:colorrate}
and \ref{fig:colorcolor}
with black diamonds. 
Sources~1--4 fall within the ULX
locus, and the next brightest sources (5--7) lie in between the ULX and
intermediate state populations.
The large degree of scatter seen in the models for the MW sources is the 
result of (a) distance uncertainties and (b) hysteresis effects 
\citep[e.g.,][]{MC03, DGK07}, which may also account for
the factor of $\sim 2$--3 offset between Sources~5--7 and the majority of MW
rates.
Also, estimates of the distance to NGC~253 itself are uncertain;
we assume a distance 3.94~Mpc \citep{Kar+03}, but other estimates
place the galaxy much closer \citep[e.g., 2.58~Mpc,][]{PC88},
which would increase the predicted rates of the MW sources in Figure~\ref{fig:colorrate}
by up to a factor of 2.
Therefore, this separation does not necessarily imply that they are ULXs,
although note that Sources~6 and 7 are considered to be ULXs by \citet{KP09}.
Alternatively, the fact that Sources~5--7 
are systematically more 
luminous than the MW BH binaries used to construct the diagnostic diagram 
could be the result of the much younger populations present in NGC~253, 
which would result in generally more luminous XRBs \citep[e.g.,][]{FLN+13}. 
Such sources, consisting of a massive BH accreting from a young massive star, 
are short lived and very rare in our Galaxy.
The color-color diagram (Fig.~\ref{fig:colorcolor}) shows their consistency with intermediate
(or, in the case of Source 6, soft) state sources as well as with ULXs.
Note, however, that a high-mass donor is not strictly necessary to produce a
high luminosity XRB \citep[see, eg.,][]{PDM04,VVN12}.

The remaining 13 sources, which fall within our diagnostic luminosity
range, are near the detection limit.
Even so, they align most with the loci of intermediate and hard state BH binaries.
The lack of soft state sources may be partially a selection effect, since the effective area
peaks in the medium band.
However, over the full band we are clearly able to detect sources down to a flux
level where sources in the soft state would be apparent, implying most of the
brightest binaries in NGC~253 are not in the soft state.
We cannot conclude more generally about the soft state population as a whole,
however, since we only consider those sources bright enough to have detectable
emission in \nustar's 4--6~keV band.
In general, we are likely catching these sources as they brighten in the hard
state and pass for the first time into the intermediate state, before they continue
into the soft state and fade.
The state of any individual source is unclear, given color uncertainties
and imperfect segregation of states on the diagrams.

The one exception to this is Source~8, which falls within the pulsar locus in
both diagrams.
Its hard spectrum, obvious from both Figures~\ref{fig:falsecolor} and \ref{fig:colorcolor},
makes it an ideal source for study with \nustar\ despite its low
4--6~keV flux, thanks to the flat/rising \nustar\ effective area up to $\sim 12$~keV.


\subsubsection{NGC~253 Spectrum}
\label{sec:results:ptsrcs:spectra}

\begin{figure}
\plotone{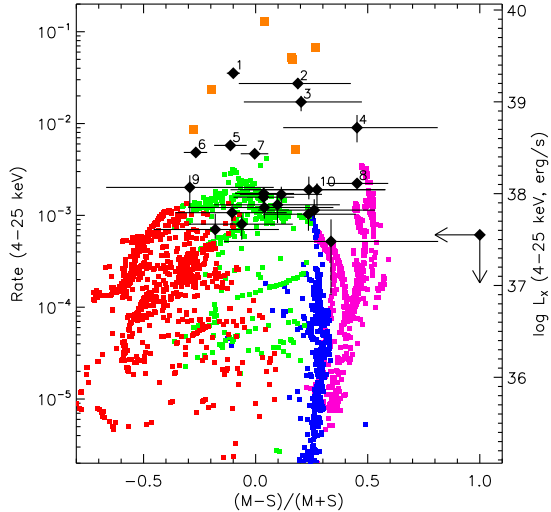}
\caption{
Hardness-intensity or ``q''-like diagram for our sources (black diamonds with 90\%
error bars or upper limits).
The hardness ratio is derived from the rates in the medium (M: 6--12~keV) and
soft (S: 4--6~keV) bands.
The 10 brightest 4--25~keV sources are labeled as in Table~\ref{tab:pts}.
Other binary types/states are depicted as squares with the following colors:
orange: ULX sources;
magenta: accreting pulsars;
blue: hard state BH XRBs;
green: intermediate state BH XRBs;
and red: soft state BH XRBs
(see Section~\ref{sec:results:ptsrcs:cmds} for details).
The ULXs are taken from other \nustar\ observations (references in the text),
while the other symbols are derived from \rxte\ observations of Milky Way 
binaries.
Count rates for these
objects and luminosities for our sources are estimated using a distance of 3.94~Mpc.
\label{fig:colorrate}}
\end{figure}

Although \nustar\ is the first observatory to resolve NGC~253 into
individual components at energies above $E \sim 10$~keV,
spatial crosstalk between many of the sources
complicates their spectral analysis.
Figure~\ref{fig:globspec} presents spectra extracted at the 
location of 6 different sources to
show their relative contribution and signal-to-noise at higher energies.
The size of the circular extraction regions for each spectrum are given in
the figure.
The ``Total Galaxy'' spectrum is extracted from a much larger aperture
(4.5\arcmin\, radius circle with the areas beyond the $D_{25}$ radius to the
northwest and southeast excluded); the larger detector area encompassed in
the region includes proportionately more background that degrades the
signal-to-noise, especially at higher energies.
Above $\sim 10$~keV, nearly all of the emission falls within 100\arcsec\, of the
galactic center and is produced almost entirely by Sources~1--4 and
Source 8 (at the highest energies).
This spectrum is extremely well fit by a broken power law with a single
Gaussian component to account for the Fe-K emission,
across the entire energy range over which counts are detected: 3--40~keV
(with $n_H$ fixed at the Galactic value).
In line with other ULX spectra, the high energy emission is soft, with steep photon indices
both below ($2.36 \pm 0.06$) and above ($3.14 \pm 0.06$) the break energy
of $5.7 \pm 0.3$~keV.
This result is consistent with the ``q''-like and color-color diagram results 
given that Source~1 and most if not all of the nuclear sources
fall within the ULX locus of Figure~\ref{fig:colorrate}.

\begin{figure}
\plotone{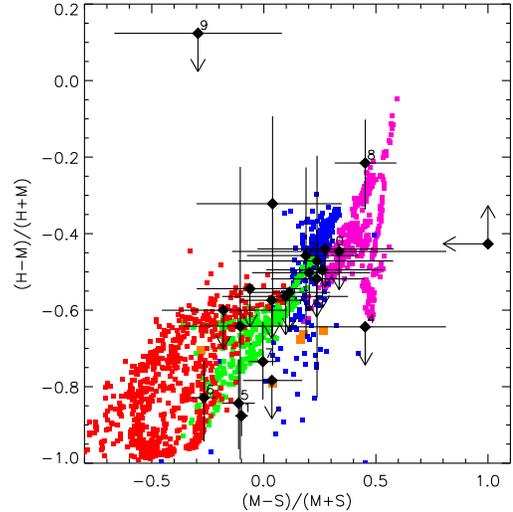}
\caption{
Color-color diagram for our \nustar\ sources.
The symbols are the same as in Figure~\ref{fig:colorrate}, and in this diagram
we also utilize the hard (H: 12--25~keV) band count rates.
Although uncertainties are large, the sources fall on the expected loci for BH XRBs.
The pulsar candidate, Source~8, is well separated from other binary accretion
modes.
\label{fig:colorcolor}}
\end{figure}

We also see that the nuclear region is much harder than many of the other sources 
and is clearly where the Fe-K line complex originates.
To first order, the overall spectrum appears dominated by a few bright sources
that are soft, with an equivalent photon index $\Gamma > 2$.

\begin{figure}
\plotone{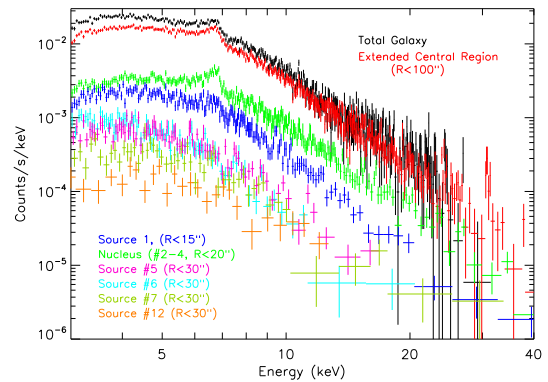}
\vspace{0.1in}
\caption{
Total NGC~253 \nustar\ spectrum (black) within $D_{25}$ relative to the spectra
of the brightest individual sources contributing to the total spectrum.
The nuclear point source emission (Sources~2--4) is shown in green and  
that from Source~1 in blue.
The emission is clearly dominated by the inner 100\arcsec of the galaxy, and 
we find that $> 99$\% of the hard X-ray flux is attributed to resolved point sources. 
The extended central region spectrum is fit well by a broken power law model with a steep
photon index of $3.14 \pm 0.06$ above $\sim 6$~keV up to 40~keV, as
described in Section~\ref{sec:results:ptsrcs:spectra}.
This region contains nearly all of the $E > 10$~keV emission from NGC~253;
the other sources contribute primarily to the spectrum at lower energies.
\label{fig:globspec}}
\end{figure}

An exception is Source~8, the pulsar candidate, which has a very flat 
spectrum ($\Gamma \sim 1$, see Section~\ref{sec:var:specfits}).
Such a hard spectrum could also be produced by a background AGN
and just happen to fall within the pulsar loci of Figures~\ref{fig:colorrate} and
\ref{fig:colorcolor}.
If so, one might expect it to have an optical counterpart.
Cursory inspection of F850LP, F606W, and F475W \hst\ ACS images
at the location of Source~8 failed to reveal any obvious counterparts.
While insufficient to rule out the classification of the source as an AGN, 
this fact does bolster the pulsar interpretation.
Because the spectra of Sources~1--4 fall off much faster above 10~keV
($\Gamma \gtrsim 3$), Source~8 makes up about 20\% of NGC~253's total
emissivity at 20~keV.  
Other fainter but similarly hard point sources
may lurk within the PSF wings of Sources~1--4 and thus go undetected.
If so, such sources might contribute significantly to the $E > 20$~keV spectrum 
of starburst galaxies generally.

\begin{figure*}
\plotone{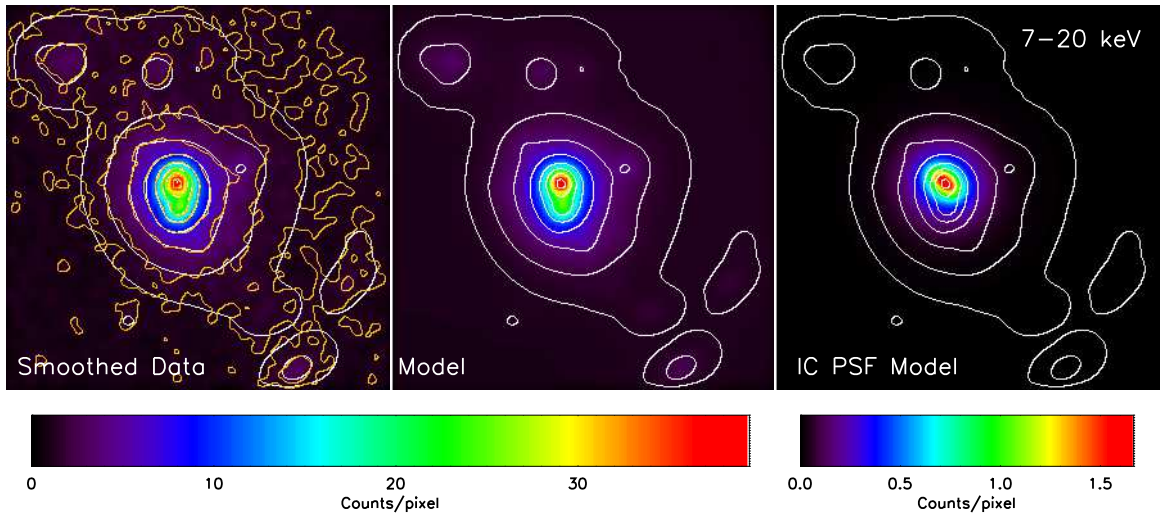}
\vspace{0.1in}
\caption{
Point source fit to the data in the 7--20~keV band used for setting limits to
the IC flux associated with the starburst (details of the left and middle panels
are the same as in Figure~\ref{fig:imgfits}).
In the right panel,
we show the PSF-convolved spatial model for IC emission,
assuming a 20\arcsec$\times$4\arcsec\, ellipse of constant surface brightness,
scaled to the expected flux in the leptonic models of \citet{LT13}.
The elliptical projected shape for the diffuse emission allows it to be distinguished 
from the bright nuclear point sources despite being co-located with them.
The leverage gained from imaging the starburst region, made possible by \nustar,
allows us to set the deepest limits on IC emission in NGC~253 to date.
\label{fig:imic}}
\end{figure*}


\subsection{Constraints on Unresolved/Diffuse Emission} 
\label{sec:results:diffuse}

There are three
likely sources of diffuse emission:  truly diffuse thermal emission,
truly diffuse non-thermal emission, and unresolved XRBs.
The thermal gas is very soft, with $kT \lesssim 1$~keV, and will contribute
only at the lowest \nustar\ energies, if at all.
Non-thermal emission is most likely to originate from cosmic-ray electrons IC scattering
the intense FIR radiation field in the starburst 
to X-ray and $\gamma$-ray energies. 
This emission should be present at some level throughout the \nustar\ band,
due to its hard ($\Gamma \sim 1.6$) spectrum.
Unresolved binaries, however, will be difficult to distinguish from the nuclear
sources given the spatial resolution of \nustar.
Otherwise they will be confused with the emission from Sources~2-4 or with
an IC component, which is assumed to have a spatial distribution similar in size 
to the starburst region.


\subsubsection{Contribution of Unresolved Point Sources and Diffuse Gas}
\label{sec:results:diffuse:pts}

An unresolved XRB population is likely brightest in the nucleus,
enhanced by HMXBs resulting from the intense star
formation there, where it is confused with
Sources~2--4.
These three sources are separated by several arcseconds, so given the large
\nustar\ PSF, a peaky spatial distribution of binaries within the central 
$\sim 75\arcsec-100$\arcsec\ would be impossible to distinguish from the
bright nuclear sources.
A slightly more extended population distributed across the entire starburst 
region or beyond could be detectable, but given the results of the next subsection
(\ref{sec:results:diffuse:ic}), we can only set upper limits on the flux of an
unresolved binary component.

The diffuse thermal gas, although soft 
\citep[$\sim 1$~keV in the hot outflow, e.g.,][]{SHW+00, MYT13}, 
may bias fits in the lowest energy bands if
no spatial model is included for its contribution.
From the 3--7~keV \chandra\ image, we construct a template surface brightness
map, excluding point sources, 
that is convolved with the \nustar\ PSF to account for its emission.
While included in fits to all energy bands, as expected this component is only
even potentially present in the 4--6~keV band; its best-fit value is $\sim 1$\%
of the combined flux of the 3 nuclear sources. 
It is not formally detected at
the 90\% confidence level.
Its morphology primarily follows the outflow to the southeast, which differs from
the other components significantly enough that we therefore expect no bias from
thermal emission in any of our results.

\begin{center}
\begin{deluxetable}{cccc}
\tablewidth{0pt}
\tablecaption{Inverse Compton 90\% Upper Limits
\label{tab:ic}}
%
\tablehead{
Projected  &  Semi-major/minor  & Upper Limit\tablenotemark{a} & 
 $\nu F_{\nu}(20~{\rm keV})\tablenotemark{b}$ \\
 Shape & axes or Radius\tablenotemark{c} & 
 \multicolumn{2}{c}{($10^{-14}$ ergs s$^{-1}$ cm$^{-2}$)}
}
\startdata
Ellipse & 20\arcsec$\times$4\arcsec & 14.2   & 16.6   \\
Ellipse & 40\arcsec$\times$8\arcsec &  11.8  &  13.8    \\
Ellipse & 60\arcsec$\times$12\arcsec &  \phn8.7  &  10.1   \\
Circle & 15\arcsec &  17.7 & 20.7    \\
Circle & 30\arcsec &  \phn6.5  &  \phn7.6  \\
Circle & 45\arcsec &   \phn3.1 &  \phn3.7  \\
Circle & 60\arcsec &   \phn2.4  & \phn2.8
\enddata
\tablenotetext{a}{Flux in the 7--20~keV band}
\tablenotetext{b}{Assuming a power law spectrum with a photon index of 1.6}
\tablenotetext{c}{1\arcsec corresponds to 19~pc at the distance of NGC~253 (3.94~Mpc)}
\end{deluxetable}
\end{center}


\subsubsection{Inverse Compton Emission}
\label{sec:results:diffuse:ic}

Because \nustar\ is the first observatory able to resolve non-nuclear 
sources away from the central starburst at $E \gtrsim 10$~keV, we have
the capability to determine whether any of the emission is both
non-thermal and diffuse.
The clean residuals for the 12--25~keV band in Figure~\ref{fig:imgfits}
already suggest that a detection of non-thermal IC emission cannot be
claimed.
However, we can place the tightest limits yet on an IC component associated
with the starburst in NGC~253, which further constrains the physical mechanisms
producing the $\gamma$-ray emission in the galaxy.

Selecting an optimal energy band for constraining the IC component requires 
maximizing the signal-to-noise ratio, where the noise is contributed by both
the background and resolved sources of emission.
Since the IC component is predicted to be relatively
hard \citep[e.g.,][]{LHB12}, we adopt a lower energy threshold of 7~keV to minimize 
soft-spectrum contributions from diffuse thermal emission
(also avoiding the Fe-K line complex around 6.5--7~keV) and individual
sources, many of which have spectral breaks near this energy. 
At the high-energy end, we encounter the relatively flat-spectrum instrumental
background and the signal-to-noise degrades.
More precisely, the background decreases with energy up to $E \sim 20$~keV, where
we encounter a complex of strong fluorescence lines.
Given these observational conditions, we restrict ourselves to the 7--20~keV
band.

Assuming IC emission originates from a disk-like region coincident with the central
starburst as in, e.g., \citet{LT13}, we expect a highly elliptical IC surface brightness
due to the large inclination of the galaxy.
This distinct appearance allows the spatial dimension to be more constraining
than the spectral dimension.
The uncertainty in the hard-band spectral indices of the 3 nuclear sources is large,
so a larger IC flux is allowed in spectral fits because the model for the point source
spectra will simply become steeper as the IC flux increases.
In contrast, spatial fits better avoid confusion between the IC and point source
components.
Figure~\ref{fig:imic} shows the point-source fit to the data in the 7--20~keV image
in the left and center panels, just as in those panels in Figure~\ref{fig:imgfits}.
We do not include the diffuse component meant to represent thermal gas
since its flux was consistent with zero in the 6--12~keV and 12--25~keV band fits.

To determine the 90\% upper limit on the IC component, we added an extended, 
PSF-convolved IC component to the best-fit
spatial model of the point-source population.  
We varied its size and intensity until C-stat
increased by an additional 2.706 above its value without the IC component.  
The right panel in Figure~\ref{fig:imic} shows a sample PSF-convolved 
IC model with an assumed 20\arcsec$\times$4\arcsec\, ellipse
of constant surface brightness.  
The total flux displayed in this spatial model is roughly consistent with the
predicted value in the leptonic models of \citet{LT13}, which amounts to 
$\sim 5$\% of the total nuclear emission.
Our upper limit for this model is $\sim 2$ times brighter.

In Table~\ref{tab:ic}, we list upper limits for a variety of simple IC geometries.
Perhaps counterintuitively, the upper limits become more stringent as
the region increases in size, despite the fact that
the IC surface brightness decreases with the size of the region
for a given flux (i.e., the same flux is spread over a larger area).
This trend is a direct result of the degeneracy between more compact diffuse
regions and the nuclear point sources.
The smaller IC regions are closer in size to the \nustar\ PSF, so that
as the diffuse IC flux is increased when deriving upper limits, 
the flux in the point sources can correspondingly
decrease to maintain a reasonable fit.
When the IC region size becomes much larger than the PSF FWHM, however,
the flux from point sources cannot compensate as well, resulting in lower flux limits
despite the fact that the IC flux is spread over a larger area.
In other words, our sensitivity to IC emission is dominated by the degeneracy
between the nuclear point source fluxes and the IC flux.


\section{X-ray and Radio Variability}
\label{sec:var}

We repeated the image analysis on each epoch individually, allowing
the detection of month-scale variability from state changes in the brightest
sources.
Considering the epochs separately also allows more physically meaningful
joint \chandra-\nustar\ spectral fits of those sources.

\begin{deluxetable*}{ccccccccc}
\tablewidth{0pt} 
\tablecaption{Per Epoch Corrected Count Rates of 
\nustar\ NGC~253 Point Sources\tablenotemark{a}
\label{tab:eprates}}
\tablehead{
 & & \multicolumn{4}{c}{\nustar\ Count Rates} & $L_X$\tablenotemark{b} & & \\
 & & \colhead{4--6~keV} & \colhead{6--12~keV} & 
 \colhead{12--25~keV} & \colhead{4--25~keV} & \colhead{4--25~keV} &
 \underline{(M-S)} & \underline{(H-M)} \\
 \colhead{ID} & \colhead{Epoch} & 
 \colhead{(10$^{-4}$ cts s$^{-1}$)} & \colhead{(10$^{-4}$ cts s$^{-1}$)} & 
 \colhead{(10$^{-4}$ cts s$^{-1}$)} & \colhead{(10$^{-4}$ cts s$^{-1}$)} &
  \colhead{($10^{38}$ erg s$^{-1}$)} & (M+S) & (H+M)}
\startdata
  & 1 & 142.0 $\pm$ 10.1 & 122.7 $\pm$ 10.9 & \phn\phn5.8 $\pm$ \phn3.8 & 274.7 $\pm$ 16.5 & 15.93 & -0.07$^{+0.05}_{-0.06}$ & $<$-1.01 \\
1 & 2 & 181.4 $\pm$ \phn9.3 & 134.1 $\pm$ 10.0 & \phn\phn7.1 $\pm$ \phn3.9 & 327.9 $\pm$ 16.7 & 19.02 & -0.15$^{+0.04}_{-0.04}$ & -0.90$^{+0.08}_{-0.08}$ \\
  & 3 & 240.2 $\pm$ 14.0 & 216.4 $\pm$ 14.4 & \phn18.5 $\pm$ \phn5.3 & 477.3 $\pm$ 21.7 & 27.68 & -0.05$^{+0.04}_{-0.04}$ & -0.84$^{+0.06}_{-0.07}$\vspace{0.2cm} \\ \hline \\
  & 1 & 110.0 $\pm$ \phn7.8 & 227.9 $\pm$ 16.2 & \phn46.4 $\pm$ \phn3.3 & 425.4 $\pm$ 30.2 & 24.68 & \phn0.35$^{+0.06}_{-0.05}$ & -0.66$^{+0.05}_{-0.07}$ \\
2+3+4 & 2 & 220.6 $\pm$ 11.3 & 368.6 $\pm$ 18.8 & \phn72.3 $\pm$ \phn3.7 & 662.9 $\pm$ 33.9 & 38.45 & \phn0.25$^{+0.04}_{-0.04}$ & -0.67$^{+0.04}_{-0.05}$ \\
  & 3 & 156.6 $\pm$ \phn9.1 & 277.1 $\pm$ 16.1 & \phn54.9 $\pm$ \phn3.2 & 496.8 $\pm$ 28.9 & 28.82 & \phn0.28$^{+0.05}_{-0.04}$ & -0.67$^{+0.04}_{-0.06}$\vspace{0.2cm} \\ \hline \\
  & 1 & \phn29.0 $\pm$ \phn3.1 & \phn20.2 $\pm$ \phn3.3 & $<$4.0 & \phn51.1 $\pm$ \phn4.6 & \phn2.96 & -0.18$^{+0.10}_{-0.09}$ & -0.82$^{+0.21}_{-0.15}$ \\
5 & 2 & \phn31.6 $\pm$ \phn3.2 & \phn26.7 $\pm$ \phn3.3 & \phn\phn3.8 $\pm$ \phn2.3 & \phn64.0 $\pm$ \phn5.8 & \phn3.71 & -0.08$^{+0.08}_{-0.08}$ & -0.75$^{+0.16}_{-0.10}$ \\
  & 3 & \phn30.0 $\pm$ \phn4.4 & \phn25.5 $\pm$ \phn4.7 & $<$1.4 & \phn56.7 $\pm$ \phn6.6 & \phn3.29 & -0.08$^{+0.12}_{-0.12}$ & $<$-1.16\vspace{0.2cm} \\ \hline \\
  & 1 & \phn32.8 $\pm$ \phn3.1 & \phn19.1 $\pm$ \phn2.7 & $<$2.7 & \phn53.3 $\pm$ \phn4.4 & \phn3.09 & -0.26$^{+0.08}_{-0.08}$ & $<$-1.07 \\
6 & 2 & \phn24.4 $\pm$ \phn2.8 & \phn12.0 $\pm$ \phn2.5 & \phn\phn2.2 $\pm$ \phn1.9 & \phn38.9 $\pm$ \phn4.2 & \phn2.26 & -0.34$^{+0.10}_{-0.10}$ & -0.70$^{+0.21}_{-0.24}$ \\
  & 3 & \phn30.5 $\pm$ \phn3.4 & \phn20.1 $\pm$ \phn3.4 & $<$3.6 & \phn55.0 $\pm$ \phn5.8 & \phn3.19 & -0.20$^{+0.10}_{-0.09}$ & $<$-1.05\vspace{0.2cm} \\ \hline \\
  & 1 & \phn30.7 $\pm$ \phn3.4 & \phn20.1 $\pm$ \phn3.3 & $<$4.0 & \phn55.2 $\pm$ \phn5.6 & \phn3.20 & -0.21$^{+0.09}_{-0.09}$ & $<$-1.02 \\
7 & 2 & \phn12.9 $\pm$ \phn2.5 & \phn20.1 $\pm$ \phn3.4 & \phn\phn5.2 $\pm$ \phn2.9 & \phn39.3 $\pm$ \phn5.0 & \phn2.28 & \phn0.22$^{+0.12}_{-0.13}$ & -0.59$^{+0.18}_{-0.17}$ \\
  & 3 & \phn18.8 $\pm$ \phn3.1 & \phn22.2 $\pm$ \phn3.3 & \phn\phn3.1 $\pm$ \phn2.5 & \phn46.1 $\pm$ \phn5.2 & \phn2.68 & \phn0.08$^{+0.11}_{-0.11}$ & -0.75$^{+0.16}_{-0.16}$\vspace{0.2cm} \\ \hline \\
  & 1 & \phn\phn6.0 $\pm$ \phn2.3 & \phn12.6 $\pm$ \phn2.9 & \phn\phn8.1 $\pm$ \phn2.3 & \phn26.8 $\pm$ \phn4.3 & \phn1.55 & \phn0.35$^{+0.20}_{-0.19}$ & -0.22$^{+0.18}_{-0.18}$ \\
8 & 2 & $<$4.0 & \phn10.9 $\pm$ \phn3.0 & \phn\phn8.0 $\pm$ \phn2.3 & \phn19.8 $\pm$ \phn4.7 & \phn1.15 & $>$0.45 & -0.16$^{+0.20}_{-0.21}$ \\
  & 3 & \phn\phn5.2 $\pm$ \phn2.7 & \phn\phn10.0 $\pm$ \phn3.2 & \phn\phn5.4 $\pm$ \phn2.4 & \phn20.1 $\pm$ \phn4.9 & \phn1.17 & \phn0.31$^{+0.28}_{-0.27}$ & -0.30$^{+0.26}_{-0.27}$ \\

\enddata 
\tablenotetext{a}{Sources' IDs are sorted by their 4-25~keV 3-epoch-summed 
count rates in descending order.}
\tablenotetext{b}{Simple conversion assuming a typical spectrum-weighted 
effective area across the band of 300 cm$^2$.}
\end{deluxetable*}

\begin{figure}
\plotone{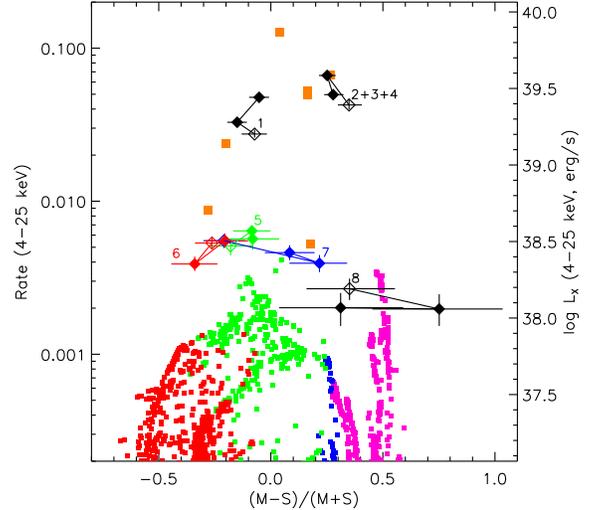}
\caption{
Hardness-intensity or ``q-like'' diagram for the brightest 8 sources in each epoch
(black and color diamonds with 90\%
error bars; upper limits are shown with error bars extending to values of 1.0).
The first epoch is indicated by an open diamond.
The symbols and band definitions are the same as in Figure~\ref{fig:colorrate}, and
the sources are labeled as in Table~\ref{tab:pts}.
While some significant variability in overall flux is seen in a few sources, only
Source~7 undergoes a significant change in hardness ratio, indicative of a
state change.
\label{fig:colorrate_ep}}
\end{figure}

\begin{figure}
\plotone{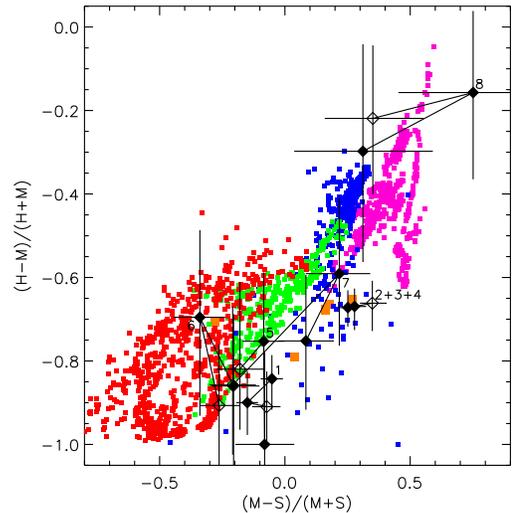}
\caption{
Color-color diagram for the brightest 8 \nustar\ sources in each epoch.
The symbols are the same as in Figure~\ref{fig:colorrate_ep},
and error bars that reach 1.0 or $-1.0$ are really upper or lower limits, respectively.
Although uncertainties are large, the ``color'' of Source~7 evolves from that
of a soft or intermediate state (red/green) to the hard state (blue).
\label{fig:colorcolor_ep}}
\end{figure}

\subsection{Image Fits}
\label{sec:var:imgfits}

The lower per-epoch depths limits us to the brightest $\sim 8$ sources
for discerning state changes between epochs.
Because the nuclear sources (2, 3, and 4) are confused in the \nustar\ data,
we previously used variability -- shown to be caused primarily by only
Source~2 -- to investigate their characteristics \citep{Leh+13}.
Given this work, we focus on the nature of the other 5 sources.

In general, each source undergoes some marginally statistically significant
variation between epochs, although largely in overall luminosity and not color.
From Epochs~1--3, Source~1 steadily increases in flux, Sources~5 and 6 exhibit slight negative
fluctuations in the second epoch, and Source~8 may have also dropped in
flux after the first epoch.
The only source to experience a clear color change is Source~7; its
4--6~keV count rate is $\sim 2$ times brighter in the first epoch than in the
other two epochs while its $\gtrsim 7$~keV emission remains unchanged.
Table~\ref{tab:eprates} gives the count rates for each source and
Figures~\ref{fig:colorrate_ep} and \ref{fig:colorcolor_ep} place these count rates
on the state diagnostic diagrams.

The hardening of Source~7 is apparent in Figure~\ref{fig:colorrate_ep},
which suggests either a transition from the soft to hard state or
oscillations between soft and intermediate states
\citep[that create the ``eye of the turtle'' in the ``q'' diagram, e.g.,][]{FBG04}.
The latter interpretation is more likely given that it occurs at higher luminosities
and that soft-to-hard transitions generally occur at $\sim 2$\% of the Eddington
luminosity \citep{Mac03}.
In Figure~\ref{fig:colorcolor_ep}, the hard band colors largely bolster
this interpretation, although the colors are also consistent with the
hard state, given the uncertainties.
Modeling the detailed spectra may be able to constrain whether the emission
is disk-dominated or not, and therefore confirm source states.
Although the results of these fits may not directly correspond to state changes
in ``q''-like diagrams \citep[e.g.,][]{DFK+10}, we nevertheless apply simple models
to our spectra in Section~\ref{sec:var:specfits}.

\subsection{Joint \chandra-\nustar\ Fits to Brightest Sources}
\label{sec:var:specfits}

Assuming variability on short (day-long) timescales is minimal, the near simultaneous
\chandra\ and \nustar\ spectra can be fit together over a broad (0.5~keV $< E <$ 25~keV)
energy range.
Narrow energy ranges can fail to discriminate between 
non-thermal and thermal-dominated spectra due to degeneracies between
highly absorbed power law and multi-color blackbody disk (MCD) spectral shapes.
Typical disk models peak in energy output around $\sim 2$--3~keV, so coverage
well beyond 3~keV is necessary to determine whether the curvature observed
below 3~keV is truly thermal and not just the result of a large absorbing column.

Because of low signal-to-noise above 10~keV, we only consider simple
non-thermal ({\tt POWERLAW}) and thermal \citep[{\tt DISKBB},][]{Mak+86} 
{\tt XSpec} spectral models,
which are fit separately in an attempt only to determine which component
dominates.
In reality, most of our spectra are a mix of the two, with some fraction of the disk
Comptonized into a non-thermal corona.
A generic and self-consistent modeling of this scenario -- convolving the disk
emission with a Comptonization model, e.g., {\tt SIMPL}$\ast${\tt DISKBB}
in {\tt XSpec} as demonstrated in \citet{SNM+09} --
unfortunately leads to unphysical results.
Even our disk-dominated sources exhibit slight excess emission above 10~keV,
but the fit pushes the composite model to complete Comptonization with a
power law component that is too steep ($\Gamma > 4$).
Because the energy range is still too low to see the non-thermal
component dominate the emission anywhere, degeneracies between absorption,
the disk innermost radius temperature, and non-thermal index produce
uninteresting results.

Due to the proximity of the sources ($\sim 1\arcmin$ separations), cross-contamination
of the \nustar\ spectra from the PSF wings of other sources is inevitable.
To counter this difficulty, we extract spectra in circular regions encompassing only 20--50\%
of the total emission (15--30\arcsec\ in radius)
and jointly fit all 8 sources with generic broken power-law
models to approximate each source's spectrum.
When a single source is later modeled in detail, the contribution of other sources
to the \nustar\ spectrum are included as a contamination model.
The contamination contribution is sub-dominant for all sources except Source~8,
which is intrinsically faint and resides nearest to Sources~1--4.

The {\tt POWERLAW} and {\tt DISKBB} best-fit parameters for each epoch and
source are given in Table~\ref{tab:topfits}.
For the soft and intermediate state sources (1, 5, and 6), the disk model generally
is a better description of the data.
The model is only really sufficient for Source~1, however; Sources~5 and 6
have moderate to significant excesses at $E > 10$~keV.
These sources are likely to be in an intermediate or possibly a steep power law state.


\begin{deluxetable*}{ccccccc}
\tablecolumns{7} 
\tablewidth{0pt} 
\tablecaption{Joint \chandra-\nustar\ Spectral Fits
\label{tab:topfits}}
\tablehead{
 &  &  &  & $n_{\rm H}$ & $\Gamma$ / $kT_{\rm in}$\tablenotemark{b} & 
 Norm\tablenotemark{c} \\
Source\tablenotemark{a} & Epoch & Model & C-stat / dof & 
($10^{22}$ cm$^{-2})$ & (- / keV) & ($10^{-3}$)\tablenotemark{d}}
\startdata
 & 1 & {\tt POWERLAW} & \phn589.7 / 473 & \phn1.06$^{+0.11}_{-0.10}$ & \phn2.67$^{+0.08}_{-0.08}$ & \phn1.68$^{+0.24}_{-0.21}$ \\
 &  & {\tt DISKBB} & \phn422.1 / 473 & \phn0.25$^{+0.05}_{-0.05}$ & \phn1.59$^{+0.06}_{-0.06}$ & 18.60$^{+3.38}_{-2.85}$ \\
1 & 2 & {\tt POWERLAW} & \phn740.8 / 503 & \phn1.37$^{+0.15}_{-0.13}$ & \phn2.71$^{+0.09}_{-0.08}$ & \phn2.39$^{+0.41}_{-0.32}$ \\
 &  & {\tt DISKBB} & \phn479.9 / 503 & \phn0.33$^{+0.06}_{-0.06}$ & \phn1.67$^{+0.06}_{-0.06}$ & 19.00$^{+3.41}_{-2.88}$ \\
 & 3 & {\tt POWERLAW} & \phn600.3 / 462 & \phn1.57$^{+0.19}_{-0.16}$ & \phn2.76$^{+0.09}_{-0.09}$ & \phn3.42$^{+0.63}_{-0.50}$ \\
 &  & {\tt DISKBB} & \phn431.2 / 462 & \phn0.34$^{+0.07}_{-0.07}$ & \phn1.76$^{+0.07}_{-0.06}$ & 18.80$^{+3.48}_{-2.93}$ \\ \hline
 & 1 & {\tt POWERLAW} & \phn336.7 / 339 & \phn0.90$^{+0.12}_{-0.11}$ & \phn2.94$^{+0.12}_{-0.11}$ & \phn0.77$^{+0.16}_{-0.13}$ \\
 &  & {\tt DISKBB} & \phn343.1 / 339 & \phn0.13$^{+0.07}_{-0.07}$ & \phn1.32$^{+0.07}_{-0.07}$ & 13.30$^{+3.96}_{-2.99}$ \\
5 & 2 & {\tt POWERLAW} & \phn352.0 / 311 & \phn0.81$^{+0.16}_{-0.13}$ & \phn2.69$^{+0.13}_{-0.12}$ & \phn0.46$^{+0.11}_{-0.08}$ \\
 &  & {\tt DISKBB} & \phn344.6 / 311 & $<$0.08 & \phn1.57$^{+0.10}_{-0.10}$ & \phn4.96$^{+1.69}_{-1.21}$ \\
 & 3 & {\tt POWERLAW} & \phn299.9 / 287 & \phn0.58$^{+0.13}_{-0.12}$ & \phn2.49$^{+0.13}_{-0.13}$ & \phn0.31$^{+0.07}_{-0.06}$ \\
 &  & {\tt DISKBB} & \phn282.4 / 287 & $<$0.07 & \phn1.55$^{+0.10}_{-0.10}$ & \phn4.99$^{+1.82}_{-1.09}$ \\ \hline
 & 1 & {\tt POWERLAW} & \phn367.1 / 329 & \phn1.22$^{+0.18}_{-0.16}$ & \phn3.32$^{+0.15}_{-0.14}$ & \phn1.39$^{+0.37}_{-0.28}$ \\
 &  & {\tt DISKBB} & \phn342.5 / 329 & \phn0.24$^{+0.09}_{-0.08}$ & \phn1.14$^{+0.06}_{-0.06}$ & 27.50$^{+8.81}_{-6.61}$ \\
6 & 2 & {\tt POWERLAW} & \phn393.7 / 319 & \phn1.01$^{+0.13}_{-0.12}$ & \phn3.21$^{+0.13}_{-0.12}$ & \phn0.93$^{+0.21}_{-0.17}$ \\
 &  & {\tt DISKBB} & \phn321.6 / 319 & \phn0.21$^{+0.08}_{-0.08}$ & \phn1.12$^{+0.06}_{-0.06}$ & 24.90$^{+7.86}_{-5.92}$ \\
 & 3 & {\tt POWERLAW} & \phn308.4 / 298 & \phn0.95$^{+0.17}_{-0.15}$ & \phn2.98$^{+0.15}_{-0.14}$ & \phn0.72$^{+0.19}_{-0.15}$ \\
 &  & {\tt DISKBB} & \phn271.9 / 298 & \phn0.16$^{+0.09}_{-0.09}$ & \phn1.25$^{+0.08}_{-0.08}$ & 15.30$^{+5.71}_{-4.12}$ \\ \hline
 & 1 & {\tt POWERLAW} & \phn361.2 / 333 & \phn0.46$^{+0.07}_{-0.07}$ & \phn2.95$^{+0.11}_{-0.10}$ & \phn0.72$^{+0.12}_{-0.10}$ \\
 &  & {\tt DISKBB} & \phn377.4 / 333 & $<$0.02 & \phn1.05$^{+0.04}_{-0.01}$ & 40.70$^{+7.76}_{-4.91}$ \\
7 & 2 & {\tt POWERLAW} & \phn275.3 / 284 & \phn0.22$^{+0.07}_{-0.07}$ & \phn2.71$^{+0.13}_{-0.13}$ & \phn0.34$^{+0.06}_{-0.05}$ \\
 &  & {\tt DISKBB} & \phn364.1 / 284 & $<$0.82 & \phn0.93$^{+0.05}_{-0.05}$ & 48.00$^{+12.10}_{-9.66}$ \\
 & 3 & {\tt POWERLAW} & \phn206.5 / 245 & \phn0.12$^{+0.08}_{-0.08}$ & \phn2.37$^{+0.13}_{-0.13}$ & \phn0.19$^{+0.04}_{-0.03}$ \\
 &  & {\tt DISKBB} & \phn328.9 / 245 & $<$0.82 & \phn1.18$^{+0.09}_{-0.08}$ & 14.00$^{+4.95}_{-3.60}$ \\ \hline
 &  &  &  &  &  & ($10^{-5}$) \\
 & 1 & {\tt POWERLAW} & \phn120.5 / 140 & \phn3.12$^{+2.51}_{-1.84}$ & \phn1.32$^{+0.45}_{-0.41}$ & \phn1.32$^{+2.08}_{-0.79}$ \\
 &  & {\tt DISKBB} & \phn123.8 / 140 & \phn1.71$^{+1.72}_{-1.23}$ & $>$3.16 & $<$1.51 \\
8 & 2 & {\tt POWERLAW} & \phn139.7 / 134 & $<$0.82 & \phn0.80$^{+0.42}_{-0.29}$ & \phn0.24$^{+0.30}_{-0.11}$ \\
 &  & {\tt DISKBB} & \phn139.9 / 134 & $<$0.60 & $>$4.49 & $<$0.45 \\
 & 3 & {\tt POWERLAW} & \phn121.0 / 115 & \phn3.54$^{+4.89}_{-2.43}$ & \phn1.38$^{+0.63}_{-0.54}$ & $<$3.80 \\
 &  & {\tt DISKBB} & \phn122.8 / 115 & \phn2.02$^{+3.20}_{-1.42}$ & $>$3.26 & $<$3.29

\enddata 
\tablenotetext{a}{Sources' IDs are given in Table~\ref{tab:pts}}.
\tablenotetext{b}{Temperature at the inner radius of the multi-color disk.}
\tablenotetext{c}{Normalization of the {\tt POWERLAW} or {\tt DISKBB} model, 
in units of photons s$^{-1}$ cm$^{-2}$ keV$^{-1}$ at 1~keV or
$[(R_{\rm in}/1~{\rm km}) / (D / 10~{\rm kpc})]^2 \cos{\theta}$, respectively, where
$R_{\rm in}$ is the innermost radius of the accretion disk, $D$ is the distance, and
$\theta$ is the inclination angle of the disk.}
\tablenotetext{d}{Scale factor for units, except for Source 8, whose values
are scaled by $10^{-5}$}
\end{deluxetable*}

\begin{figure}
\plotone{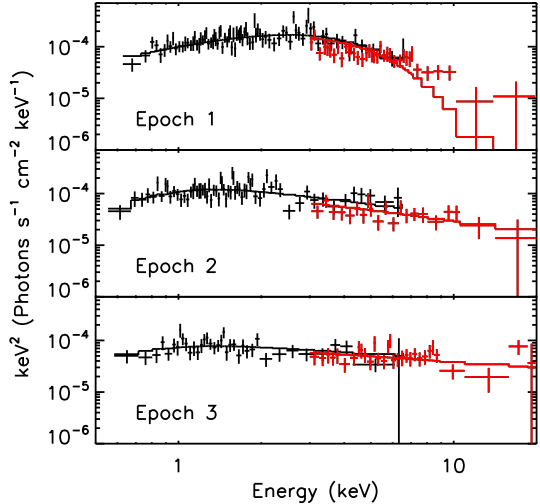}
\caption{
Source~7
\chandra\ and \nustar\ spectra fit to the {\tt DISKBB} (Epoch~1) or {\tt POWERLAW}
(Epochs~2 and 3) models.
Over the course of the observations, the spectrum hardens, primarily due to a
loss of flux below $\sim 8$~keV, likely the result of a diminishing disk component.
\label{fig:src7spec}}
\end{figure}

Source~7, while statistically preferring the non-thermal model, is better
described as becoming more non-thermal over the course of
the observations (Fig.~\ref{fig:src7spec}).
During the first epoch, its spectrum looks much like that of Source~5, consistent
with a highly absorbed steep power law  ($\Gamma = 3.0 \pm 0.1$), 
where the model parameters mimic
a hybrid thermal/non-thermal shape and do not represent true physical conditions
in the system.
In the subsequent epochs, the spectrum hardens and the disk contribution generally
diminishes, as evidenced by that hardening and the falling value of $n_{\rm H}$.
The application of more complicated models would be necessary to physically
interpret the transition, 
but this is not warranted by the signal-to-noise of the spectra.
However, the spectral fits provide further evidence that Source~7
is transitioning to the intermediate state.
Its Epoch~3, 2--7~keV \chandra\ count rate is approaching 
the lowest value measured across all archival \chandra\
observations since 2000, consistent with soft-to-intermediate
state movement on the upper left part of the ``q''-like diagram \citep{FBG04}.

Source~8, unlike all of the other sources, clearly has a hard spectrum.
Although both models appear to describe the spectra almost equally well, 
the disk inner radius temperature would have to be atypically high.
The hard ($\Gamma \sim 1$) spectrum is consistent with other accreting
pulsars in outburst \citep[e.g.,][]{Miy+13}, but
the source is too faint to see the typical high energy ($E \gtrsim 20$~keV) curvature
if it is an accreting pulsar.
Archival \chandra\ data reveal that Source~8 is roughly persistent, exhibiting little to no
variability between the 5 observations over 12 years in which it could have been detected.
The lack of variability argues against it being a transient Be/XRB, unless
it is continually outbursting.

We also investigated the long term \chandra\ variability of all of our sources.
In general, the brighter sources exhibit some variation in their
2--7~keV count rates, while 
fainter sources lack photon statistics necessary for variability constraints.
Only two sources (15 and 18) are clear transients, having been detected for the
first time in these observations.
Source~15 is detected in Epoch~3 alone, and Source~18 is undetected in
the first epoch but is growing in flux from Epoch~2 to 3.
The uncertainty in their \nustar\ measurements, however, precludes us from
concluding anything about their nature based on hard energy data.

\subsection{Radio Monitoring}
\label{sec:var:radio}

The VLBA campaign was intended to catch flares of similar intensity to those
observed in Cyg X-3.
In individual epochs, no flares were detected above our 
rms ($1\sigma$) noise of $\sim$150\,$\mu$Jy\,beam$^{-1}$.

Within the core of NGC~253, we detect the two brightest known VLBI SN remnants,
but no new sources were detected.
This is not surprising since:
(1) most radio sources in the cores of starburst galaxies are diffuse H{\sc ii} regions or 
supernova remnants 
\citep[e.g., in M82:][]{MMB+02,GFB+13}
and at typical expansion speeds of 10,000\,km\,s$^{-1}$ would be resolved out by 
the VLBA after 300 years; 
(2) there is significant free-free absorption towards 
the core of NGC~253 
\citep[e.g.,][]{Tin04,LT06,RML+14};
(3) the predicted supernova rate is low \citep[$<0.2$\,yr$^{-1}$;][]{RML+14}; 
and (4) there is no radio evidence for an AGN \citep{Bru+09}.  
In the wider galaxy, the lack of detections corresponding to the 
\chandra\ and \nustar\ sources is also not surprising given the probability
of catching a flare, but we are able to put a limit on
their radio brightness at the time of observation.  
For phase centers correlated in all three epochs, the rms in the images made 
from combining all three epochs is $\sim$65\,$\mu$Jy\,beam$^{-1}$.  
At a distance of 3.94~Mpc, this puts a $5\sigma$ limit of 
$6\times10^{17}$\,W\,Hz$^{-1}$ on the brightness of individual counterparts.


\section{Discussion}
\label{sec:disc}


\subsection{Extragalactic Point Sources at Hard Energies}
\label{sec:disc:ptsrcs}

We present the first imaging observations above 10 keV for a galaxy outside the 
Local Group. 
For the first time, we are able to spatially resolve the $> 10$~keV X-ray emission 
of NGC~253 into individual sources, revealing that the galaxy's overall spectrum
turns over (is relatively X-ray soft) above 10 keV and is dominated by a small 
number of luminous sources that also show turnovers above 10 keV. 
Source rates and colors (i.e., hardness ratios) are used to characterize source 
types through diagnostic plots, in which the spectra of MW BH binaries have been 
translated into NuSTAR rates and colors.

Comparison of MW binaries to our sources suggests that the majority (by number) 
of the NGC~253 XRB population are black holes primarily in the intermediate and 
possibly hard state, which is dominated by a power-law/non-thermal component.
Since observations of external galaxies give us a view of the entire galaxy
simultaneously, we can effectively constrain the dominant states
of all binaries at several snapshots in time.
Direct comparison of the near-Eddington accreting sources we detect
in the MW is problematic;
as is clear from Figure~\ref{fig:colorrate}, we hit our detection
threshold roughly where we expect the brightest MW binaries to be.
Only a single XRB in the Milky Way, GRS~1915+105, has spent
long periods of time with an X-ray luminosity at or near the Eddington
luminosity.  
\citet{RBV03} examined a large number of
observations, and concluded that GRS~1915+105 is nearly always in the
``very high state,'' consistent with the bright
intermediate state.
Some sources may be expected to be caught in
extremely bright hard states as transients \citep[e.g., V404 Cyg,][]{Oos+98}.

We also find one source, 8, that may be an accreting pulsar
based on its position in the hardness/intensity diagrams
(Figs.~\ref{fig:colorrate} and \ref{fig:colorcolor}). 
Given the short duration of the type-I outbursts (associated with neutron stars)
of Be XRBs,
and the very rare occurrence of the more energetic type-II outbursts
\citep[e.g.,][]{Rei11}, we would not expect a very large number of
these systems in the few snapshots we have obtained.
Source~8, however, is persistent, having been detected in all sufficiently
sensitive \chandra\ observations at roughly the same flux.
This persistence suggests we have not observed a single long or multiple
$L_X \sim 10^{38}$~erg s$^{-1}$ outbursts,
but instead an extremely X-ray luminous pulsar.
Because the source is faint and seemingly heavily absorbed, 
the intrinsic spectrum may not be as hard as observed.
In this case, the actual accreting object may be a stellar-mass BH XRB or 
an obscured AGN behind the galaxy.

Interestingly, the most luminous \nustar\ sources in NGC~253 \citep[including
the most luminous source in the neighborhood of the nucleus;][]{Leh+13}  
are located in the region of color-intensity (``q''-like) and color-color diagrams
occupied by \nustar-observed ULXs.
Sources~6 and 7 have in fact been considered ULXs in a previous study \citep{KP09}.
Unlike Source~1, a clear ULX in all of our observations, the spectra of
Sources~5--7 favor a hard non-thermal component in addition to the thermal component,
which strongly dominates \nustar-observed ULXs 
\citep[e.g.,][]{Bac+13,Wal+13,Ran+14,Wal+14}.
Their variability over the past 12~years, however, suggests they may very well 
exhibit ULX-like luminosities, even if they appear as borderline ULX candidates
in these observations.

\begin{figure*}
\begin{center}
\includegraphics[width=6.8cm, angle=270]{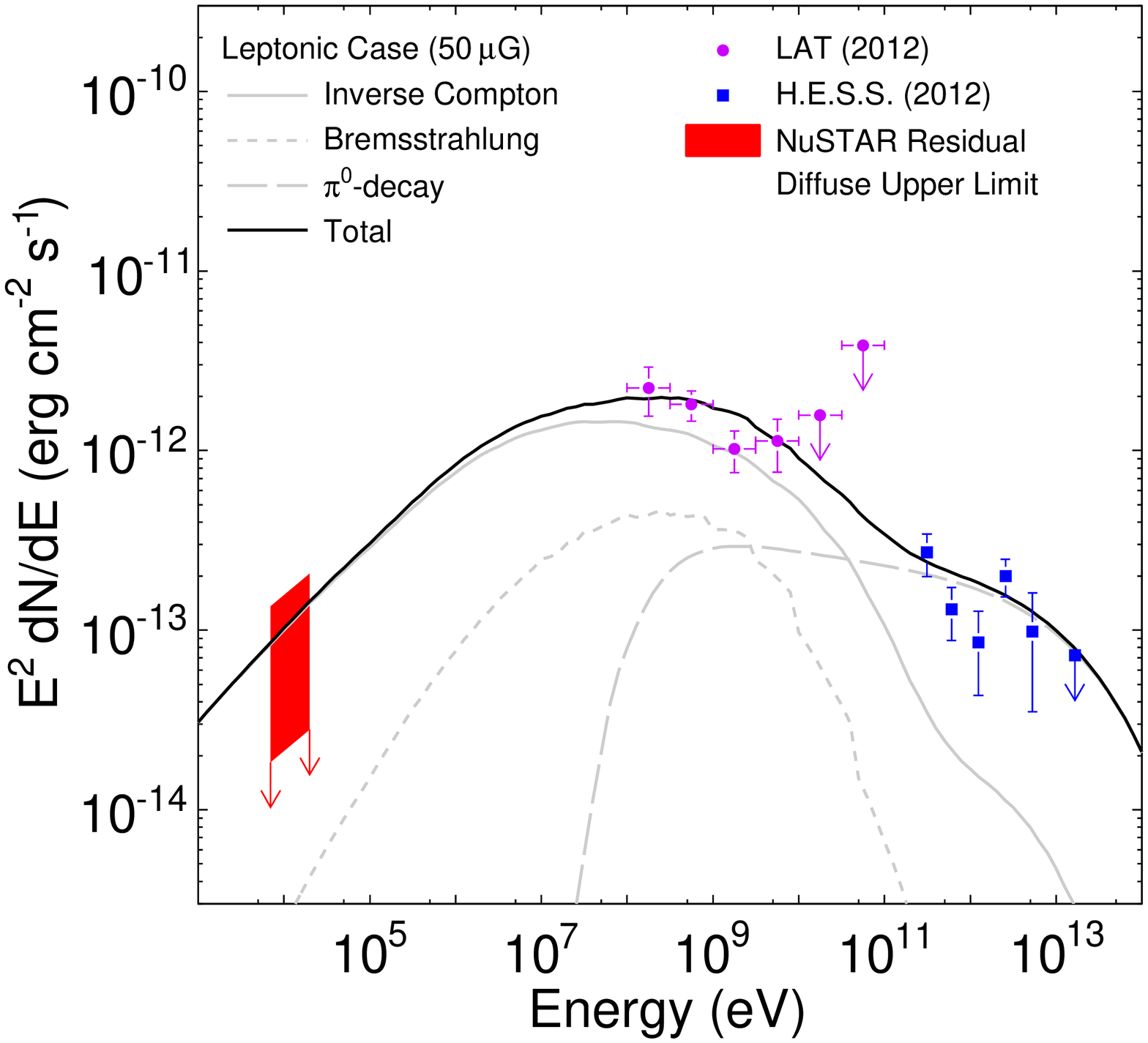}
\includegraphics[width=6.8cm, angle=270]{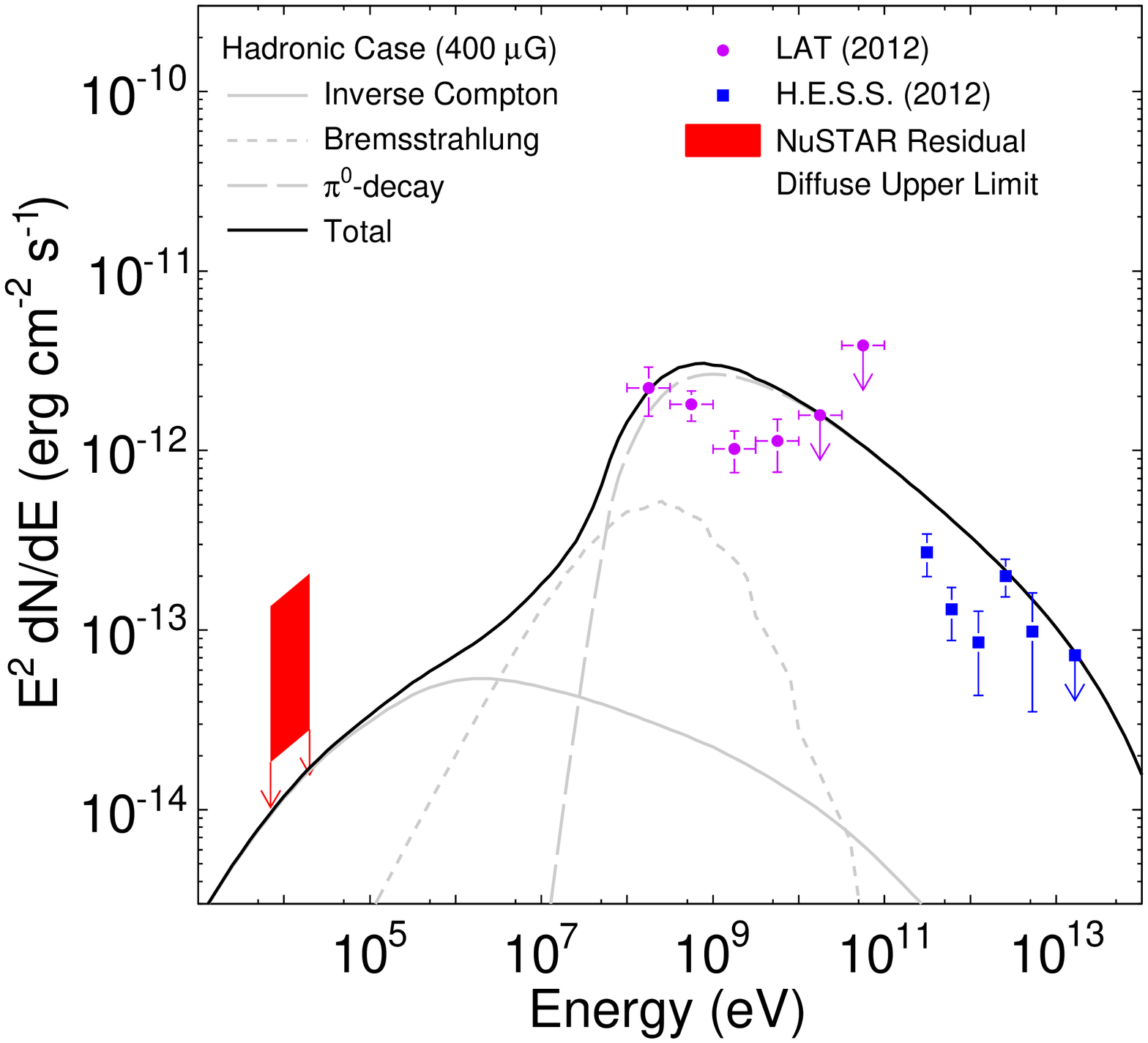}
\vspace{0.1in}
\caption{
Broadband (X-ray to $\gamma$-ray) modeling of cosmic ray emission mechanisms
under leptonic (left panel) and hadronic (right panel) scenarios,
using the model parameters of \citet{LHB12}, which are based on the
the GeV \citep[\fermi\ LAT,][]{Ack+12}
and TeV \citep[H.E.S.S.,][]{Abr+12} detections of NGC~253.
The range of upper limits at X-ray energies provided by \nustar\ 
are given by the red/shaded box (see Table~\ref{tab:ic} for numerical
values and corresponding emission region sizes).
Although typical leptonic models are ruled out for large IC emitting regions,
small regions confined to the starburst itself are allowed by these 
\nustar\ constraints.
\label{fig:sedfit}}
\end{center}
\end{figure*}


\subsection{Constraints on Non-thermal Emission}
\label{sec:disc:ic}

The spatial resolution and effective area at $E > 10$~keV provided by
\nustar\ has allowed the most sensitive constraint on IC emission in a starburst
galaxy to date.
In Section~\ref{sec:results:diffuse:ic}, we derive upper limits for various
assumptions of the spatial distributions of the IC-emission.
Although we are not quite able to use the upper limits to discriminate
between the leptonic and hadronic scenarios that can both describe
the $\gamma$-ray emission from NGC~253, we consider each scenario
in comparison to our results.

The evolution of cosmic-ray nuclei and electrons is determined by the 
diffusion-loss equation \citep[see, e.g.,][]{Lon94}:
\begin{equation}\label{eqn:diff-loss}
-D\nabla^2N(E)+\frac{N(E)}{\tau(E)}-\frac{d}{dE}\left[b(E)N(E)\right] - Q(E) = -\frac{\partial N(E)}{\partial t} \,,
\end{equation}
where $D$ is the scalar diffusion coefficient, $\tau(E)$ is the timescale for particles 
with energy $E$ to escape the region, $b(E)$ is the cooling rate for the particles, 
$Q(E)$ is the source term, and $N(E)$ is the number density of particles with 
energies in the range $E$ and $E+dE$. 
In our modeling, we assume that the system is in steady state 
($\partial N(E)/\partial t = 0$) and that the spatial dependence of the diffusion term 
can be neglected ($D\nabla^2N(E) = 0$). 
Equation~\ref{eqn:diff-loss} can be solved using the Green's function \citep{TRD+04}
\begin{equation}
G(E,E')=\frac{1}{b(E)}\exp\left(-\int_{E}^{E'}dy\frac{1}{\tau(y)b(y)}\right)\,,
\end{equation}
which for a given source term, $Q(E)$, results in the steady-state solution given by
\begin{equation}
N(E) = \int_{E}^{E_{\rm max}}dE'Q(E)G(E,E')\,.
\end{equation}
From this particle distribution, we can compute the broadband non-thermal 
diffuse emission by convolving with the spectra of the various cooling interactions 
and the target particles. 

In the case of cosmic-ray nuclei, the non-thermal diffuse emission predominantly at 
$E \gtrsim100$~MeV arises from pion production interactions with interstellar gas.
In the case of cosmic-ray electrons and positrons, the emission extends from X-rays 
through GeV $\gamma$-rays and arises from bremsstrahlung interactions with 
interstellar gas and IC scattering of interstellar radiation. 
These interactions are also included in the cooling rates, $b(E)$, for the various 
particle species. 
We also account for cooling due to ionization (for all cosmic-ray species) and 
synchrotron (electrons and positrons). 
Additional losses due to particle escape ($\tau(E)$) via diffusion and/or 
advection due to starburst winds are included in the model. 
For positrons, annihilation is included as an additional escape term. 
Primary particles ($Q(E)$) are assumed to be accelerated in supernova remnants, 
and are injected with power-law spectra 
\citep[$\Gamma \sim 2.1$; see][]{LHB12,CF13}.
Secondary electrons and positrons from pion-production 
\citep[computed using the analytical formulae from][]{KAB06}
and from ionization by cosmic-ray nuclei are included in the electron source term for 
computation of the final electron/positron distributions. 
The final broadband diffuse spectrum is calculated assuming the best-fit physical 
parameters (i.e., supernova rate, acceleration efficiencies, galaxy gas mass, 
starburst wind speed, diffusion timescales, and region sizes) for the 
leptonic ($B \sim 50\mu$G) and hadronic ($B \sim 400\mu$G) models in 
\citet[][see their Table~1]{LHB12}. 
However, for the interstellar radiation field (the seed photons for the IC emission), 
we adopt the radiative transfer model from \citet{SK07}.
Viable diffuse models are required to reproduce both the {\it Fermi}-LAT 
\citep{Ack+12} and H.E.S.S.\ \citep{Abr+12} data points.

In Figure~\ref{fig:sedfit}, we plot two models (one in each panel) 
for the $\gamma$-ray emission in NGC~253 under scenarios in which the emission
is dominated either by leptonic or hadronic processes.
To match the GeV and TeV observations, 
different assumptions of the cosmic-ray density and magnetic field strength
are made in each case. 
The leptonic model (left panel) results in much more
non-thermal emission in the hard X-ray band than in the hadronic model (right panel). 
In both panels, the \nustar\ upper limits on the broadband diffuse component
for various assumptions about the size of the emission region
are given by the shaded region. 
As the dense gas and radiation environment of the starburst is expected to 
prevent electrons from diffusing too far in the disk of the galaxy, the size of the 
emission region is expected to be roughly the size of the starburst core 
($R \sim 350$ pc), 
which roughly corresponds to an angular size of $\sim 20\arcsec$. 
The corresponding \nustar\ upper limits -- the upper part of the shaded bands
in Figure~\ref{fig:sedfit} -- are comparable to the hard X-ray
diffuse emission in the leptonic scenario, 
although larger emission region size estimates yield more stringent constraints. 
Hadronic models, on the other hand, have substantially less hard X-ray emission, 
and thus are out of reach for even the most optimistic size estimates.

Further modeling efforts regarding the spatial and spectral properties of
the IC component (beyond the scope of this work) are needed to fully
interpret the \nustar\ observations of NGC~253. 
Population synthesis
models to account for individual sources too faint to be
individually detected in the \nustar\ band would also enable more sensitive
constraints on the IC emission; XRBs are expected to dominate 
the emission in low redshift star-forming galaxies \citep{Leh+10,SSK14}. 
Still, the present constraints
generally disfavor scenarios in which the $\gamma$-ray luminosity is attributed
primarily to leptonic processes, providing further support for
enhanced cosmic-ray energy density associated with actively star-forming environments.


\subsection{The Global 0.5-40~keV Spectrum}
\label{sec:disc:globspec}

To place these \nustar\ results a broader context, 
we construct and model \chandra\ and \nustar\ spectra for the entire NGC~253 galaxy.
For the \chandra\ response files, RMFs and ARFs are weighted by the spatial
distribution of emission, an approximation that works well given its concentrated PSF.
The \nustar\ emission, entirely made up of what can effectively be considered point sources,
is much less localized due to the larger PSF, making it inaccurate to simply weight the
response files by the emission distribution in the same way.
Instead, we assume all of the emission originates from the positions of the sources
in Table~\ref{tab:ic}, weighted by their relative 4--25~keV count rates, to construct an
average ARF for use with the global spectrum.

While the contributors to the \nustar\ spectrum are effectively point-like, at lower
energies thermal gas quickly dominates the global X-ray spectrum.
Using the unresolved \chandra\ emission as a guide, we model it
as a three-temperature plasma representing disk (cooler) and wind (hotter) gas:
(1) an unabsorbed 0.3~keV component representing higher radius disk and halo
gas, (2) a moderately absorbed 0.6~keV component representing warmer disk
and the large-scale wind emission, and (3) a highly absorbed 2~keV component
representing superwind emission associated with the nuclear starburst.
The latter component is confined to within the nuclear starburst region, whereas
the cooler gas components are much more extended.
The remaining detected emission is entirely from point sources, which we model
as a broken power law with best-fitting indices of 1.5 and 3 below and above,
respectively, the break energy of 6.2~keV.
For each of these components, we assume solar abundances.
This model is obviously not physical, but successfully acts as an average
representation of emission from multiple disk-blackbody and power law spectra 
with a variety of temperatures and indices.
In Figure~\ref{fig:globspec}, the unfolded energy spectrum illustrates the relative
contributions of these components in the 0.5--40~keV band.
We also insert an IC component (with a photon index of 1.6) pegged at our most
conservative upper limit from Table~\ref{tab:ic} to show its relative,
maximal importance at these energies.

\begin{figure*}
\begin{center}
\includegraphics[width=5in]{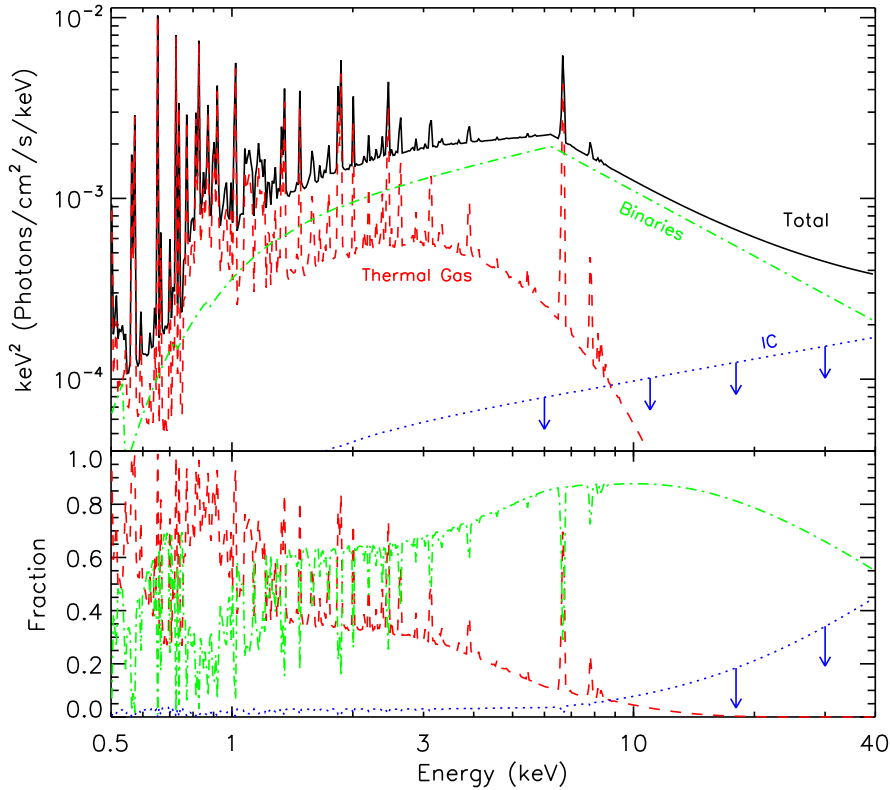}
\vspace{0.1in}
\caption{
Unfolded model of the X-ray emission from NGC~253 based on fits to global
\chandra\ and \nustar\ spectra over the 0.5--40~keV energy range (top panel).
The fit includes three {\tt APEC} models representing
thermal gas from the disk and starburst-driven winds (``Thermal Gas,'' red/dashed line), 
a broken power law model
incorporating all the emission from XRB point sources (``Binaries,'' green/dot-dashed line), 
and a power law model
indicating our most conservative upper limit for IC emission as found in 
Section~\ref{sec:results:diffuse:ic} (``IC," blue/dotted line).
The fraction of emission attributed to each component is given in the bottom panel.
Point source emission prevails above $\sim 1.5$~keV and peaks between
6 and 7~keV, declining at higher energies due to the intermediate-like states of
the ULX sources that dominate the \nustar\ spectrum.
Diffuse IC emission does not contribute appreciably below 40~keV.
\label{fig:globspec}}
\end{center}
\end{figure*}

Although the observed luminosity is not a strong function of energy, 
it peaks in the 2--10~keV band
with $L_X = 7.3 \times 10^{39}$~\lum, compared to the slightly lower luminosities
at lower ($L_X = 4.1 \times 10^{39}$~\lum, 0.5--2~keV) and higher
($L_X = 2.1 \times 10^{39}$~\lum, 10--40~keV) energies.

\subsection{Contribution of Starburst Galaxies to the CXB}
\label{sec:disc:spectrum}

The CXB peaks in $\nu F_\nu$ at $E \sim 30$~keV \citep[e.g.,][]{GMP+99} and
has yet to be fully resolved into contributing source populations at $E > 10$~keV.  
Focusing hard X-ray telescopes, of which \nustar\ is the first, are expected to make 
major headway on this issue and it is expected that up to 50\% of the hard CXB will 
ultimately be 
resolved in \nustar\ deep surveys \citep{BDM+11}.   
While it is clearly the case that AGN and clusters dominate the overall flux of the 
CXB at energies below 10 keV \citep[e.g.,][]{WFB+06}, 
starburst galaxies, given their large numbers and strong evolution with cosmic time 
(there are many more luminous starburst galaxies at high redshift) could have a 
non-negligible contribution to the hard CXB.  
This idea was put forth in \citet{PR03}, who took a template X-ray spectrum for 
starburst galaxies and calculated their contribution to the 
CXB assuming that their density evolves as 
$(1+z)^q$ up to  $z = 5$.  
They found that at energies $E \lesssim 15$~keV this contribution is at a 
level of a few percent for $q \le 3$.  
Recent deep \chandra\ surveys have found {\it luminosity} evolution consistent 
with lower values of $q$  
\citep[$q = 2-3$; e.g.,][]{Nor+04,PMH+07,TG08,Tre+13}.   
However,  \citet{PR03} also predicted that the IC component 
(see Section~\ref{sec:disc:ic}) would be the main contributor to starburst galaxy emission at 
$E>10$~keV and that its relative contribution would get progressively 
higher for increasing redshift.

We thus compare the \nustar\ NGC~253 spectrum, which we have modeled 
extensively in Section~\ref{sec:results:ptsrcs:spectra}, 
to the model of \citet{PR03} to determine 
what the possible implications may be for the contribution of starburst galaxies 
to the CXB.   
Their model consists of (i) an unabsorbed 
0.8 keV thermal bremsstrahlung component from diffuse gas, 
(ii) an exponentially cutoff power-law representing the XRB populations, 
with photon index $\Gamma = 1.2$ and cutoff 
energy of 7.5~keV absorbed through $n_{\rm H}$ = $10^{22}$ cm$^{-2}$,
(iii) a similarly absorbed 
power-law with photon index $\Gamma_{1} = 1.8$ representing
the IC emission upscattered from the FIR, and 
(iv) a very faint unabsorbed power-law with photon index $\Gamma_{2} = 2.3$ 
representing the IC emission upscattered from the cosmic microwave background.   
The model of \citet{PR03} estimate that the latter two components, (iii) and (iv), 
respectively account for
5\% and 0.5\% of the 2-10~keV flux of starburst galaxies,
and $\sim 10$\% of the flux at 20~keV.

In comparison to the \nustar\ spectrum of NGC~253, we find that the 
cutoff power-law for the XRB population is too flat.  
Even with the cut-off, we require a
power law slope of $\Gamma > 2$.
Note that if we lacked the \nustar\ spatial resolution and applied the 
model of \citet{PR03} to the full NGC~253 hard X-ray spectrum, 
the contribution of Source~8 may have been interpreted as a (weak) IC component.
If the source is an accreting pulsar, we expect the spectrum to turn over quickly
above $\sim 20$~keV, so misidentifying it with IC emission would lead to
incorrect conclusions for NGC~253's output at energies above 20~keV.
If it is instead a background AGN, then the spectrum of the galaxy would of course
be even softer at hard energies. 
Our constraints on the IC component  
show that likely $<1$\%  of the 2--10~keV flux and $<2$\% of the 10--30~keV flux, 
from starburst galaxies arises from IC emission.
Importantly, the overall {\it normalization} of the 10--30 keV flux is much lower 
than previously assumed.

Detailed modeling of the type conducted in \citet{PR03} is beyond 
the scope of this paper. 
The models shown in their Figure~3, however, allow one to determine resolved 
fractions based on values for $q$ (see above) and also for the contribution by 
XRB populations and IC emission.
Choosing $q=3$ and neglecting the IC components (iii) and (iv),
which we find to be much lower and likely insignificant, we arrive at a starburst 
galaxy contribution to the $E > 10$~keV CXB of $<1$\%.


\subsection{Variability}
\label{sec:disc:var}

\subsubsection{X-ray Fluxes}
\label{sec:disc:var:xray}

Due to sensitivity limitations, only 8~sources are sufficiently bright to investigate
flux variations among epochs.
Of these, only Source~7 undergoes a clear state transition. 
Source~1 varies solely in overall luminosity, Sources~5, 6, and 8 show no 
significant variations,
and the three nuclear sources were addressed in \citet{Leh+13}.
\nustar\ separates the hard emission of these
sources -- except the nuclear ones -- and allows the identification of black
hole and neutron star binary states more robustly than is possible otherwise.

Distinguishing one absorbed spectral model from another in lower signal-to-noise 
data at $E \lesssim 10$~keV can be challenging, since only at hard energies do
degeneracies caused by the effect of absorption vanish.
\nustar's collecting area near to and just above 10~keV provides a stronger
lever arm to distinguish true MCD components from highly absorbed
power law spectra.
Although the soft/intermediate states of Sources~5-7 are in general better fit by
a pure MCD model, they also exhibit slight hard energy excesses above that
model that are even better fit by the addition of a non-thermal component.
We cannot place strong constraints at $E > 20$~keV to perform fits similar to 
binaries in the MW \citep[e.g.,][]{SNM+09,Nat+14}. 
Even so, we are able to rule out simple power law descriptions of the data, confirming
much of the emission originates from a disk-like component in these XRBs.

\subsubsection{VLBA Flare Monitoring}
\label{sec:disc:var:radio}

Among the Galactic XRBs, the ones with the strongest radio emission, 
Cyg~X-3 and SS~433, have high mass donors.  
Presumably this is because the jets from these systems interact with the 
winds of the mass donors, leading to more efficient dissipation of energy, 
and hence a higher radiative efficiency for the jet.  
Cyg X-3 shows several multi-Jy flares per year \citep[e.g.,][]{Wal+95}, 
so if the rate of such flares scales with the star formation rate of the host galaxy, 
we might expect a $\gtrsim 100$~$\mu$Jy flare every few days in NGC~253.
The span of our three 8~hr exposures as originally conceived allowed for an excellent
chance of catching such a flare.
Our probability of detecting a flare was, however, diminished by lower-than-expected
sensitivity ($\sim 150$~$\mu$Jy~beam$^{-1}$)
due to telescope failures or missing data, interference, 
and the low declination of the galaxy.
No flares were detected.

There have already been examples of radio/X-ray monitoring, e.g., in M82, 
that have turned up extremely luminous radio flares. 
One example is the recent detection of a faint radio source in the nuclear region of M82 
using MERLIN by \citet{Mux+10}. 
The true nature of this source remains unclear but it may be the first detection 
of radio emission from an extragalactic microquasar.  
Recent \chandra\ analysis by \citet{JMF11} shows that the source is not especially 
bright in X-rays, and hence that it may be an extra-bright SS~433-like object, 
rather than an extremely radio-bright ``normal'' microquasar.  
Radio emission has also been detected from an ULX in M31 \citep{Mid+13}, 
albeit at a radio flux well below what could be detected at the distance of the 
nearest starburst galaxies.


\subsection{Future Work}
\label{sec:disc:future}

NGC~253 is the benchmark deep observation in a sample of starburst galaxies
that have or will have concurrent observations with \chandra\ and \nustar.
These galaxies are listed in \cite{Har+13} and together
provide an in-depth view of XRB populations in the hard X-ray band over
a range of stellar masses and star formation rates.   
This program is critical for
understanding the ionizing output of XRB populations and of particular
interest is the much deeper view of high-redshift galaxies coming up later
in 2014 via the {\it Chandra} Deep Field 7~Ms survey (P.I. Niel Brandt), which
should detect star-forming galaxies at $z\gtrsim 4$ \citep{Bas+13}.
Note that the observed 2--10~keV {\it Chandra} bandpass corresponds
to restframe $E = 10$--50~keV at $z=4$.  
Given that XRB populations in starburst galaxies may
rival AGN as an ionizing source during the critical reionization
period of the Universe \citep{FLN+13} -- but that the bolometric correction
from total starburst luminosity to X-ray bandpass depends sensitively on
the spectrum in the hard X-ray bandpass -- this \nustar\ program has an important
role to play.

We also note that {\it Astro-H} will launch late in 2015 and will contain a hard X-ray 
instrument, the Hard X-ray Imager (HXI), with a bandpass similar to that of \nustar.   
The HXI has a slightly larger PSF than \nustar\ at $\sim 1.8$\arcmin; 
however, the background is anticipated to be slightly lower and the collecting area higher, 
so the overall sensitivity should be comparable for overall detection of NGC~253 and other
starburst galaxies.  
We expect that multiple observations from both {\it Astro-H} and \nustar, 
collected over the years, will provide highly valuable constraints on accretion state 
transitions of the bright X-ray binary population in NGC~253.  
The accumulated broad-band spectrum from these observations would substantially 
improve measurements of the $E > 30$~keV emission and the corresponding 
contribution from IC emission.

\acknowledgments
We thank the referee for insightful suggestions that improved the paper.
This research was supported by an appointment (DRW) to the NASA 
Postdoctoral Program at the Goddard Space Flight Center, administered 
by Oak Ridge Associated Universities through a contract with the
National Aeronautics and Space Administration (NASA)
and made use of data from the \nustar\ mission, a project led by the 
California Institute of Technology, managed by the Jet Propulsion Laboratory, 
and funded by NASA. 
We thank the \nustar\ Operations, Software and Calibration teams for support 
with the execution and analysis of these observations. 
This research has made use of the \nustar\ Data Analysis Software 
(NuSTARDAS) jointly developed by the ASI Science Data Center (ASDC, Italy) 
and the California Institute of Technology (USA). 
This work was also funded by a \chandra\ grant for 
Program \#13620679 (P.I. Hornschemeier).
The National Radio Astronomy Observatory is a facility of
the National Science Foundation operated under cooperative agreement
by Associated Universities, Inc.  
This work made use of the Swinburne
University of Technology software correlator, developed as part of the
Australian Major National Research Facilities Programme and operated
under license.


\end{document}